\documentstyle[aps,eqsecnum,preprint,floats,epsf,psfig]{revtex}
\begin{document}
\def\be{\begin{eqnarray}}
\def\en{\end{eqnarray}}
\def\non{\nonumber}
\def\la{\langle}
\def\ra{\rangle}
\def\nc{N_c^{\rm eff}}
\def\vp{\varepsilon}
\def\vma{{_{V-A}}}
\def\vpa{{_{V+A}}}
\def\m{\hat{m}}
\def\ov{\overline}
\def\etapp{\eta^{(')}}
\def\fp{{f_{\eta'}^{(\bar cc)}}}
\def\half{{{1\over 2}}}
\def\pr{{\sl Phys. Rev.}~}
\def\prl{{\sl Phys. Rev. Lett.}~}
\def\pl{{\sl Phys. Lett.}~}
\def\np{{\sl Nucl. Phys.}~}
\def\zp{{\sl Z. Phys.}~}
\def\lsim{ {\ \lower-1.2pt\vbox{\hbox{\rlap{$<$}\lower5pt\vbox{\hbox{$\sim$}
}}}\ } }
\def\gsim{ {\ \lower-1.2pt\vbox{\hbox{\rlap{$>$}\lower5pt\vbox{\hbox{$\sim$}
}}}\ } }

\font\el=cmbx10 scaled \magstep2
{\obeylines
\hfill 
\hfill September, 1998}

\vskip 1.5 cm

\centerline{\large\bf Charmless Hadronic Two-body Decays of $B_s$ Mesons}
\medskip
\bigskip
\medskip
\centerline{\bf Yaw-Hwang Chen,$^a$~ Hai-Yang Cheng,$^b$~ B. Tseng$^{b}$}
\medskip
\bigskip
\centerline{$^a$ Department of Physics, National Cheng-Kung University}
\centerline{Tainan, Taiwan 700, Republic of China}
\medskip
\centerline{$^b$ Institute of Physics, Academia Sinica}
\centerline{Taipei, Taiwan 115, Republic of China}
\bigskip
\bigskip
\bigskip
\centerline{\bf Abstract}
\bigskip
{\small 
Two-body charmless nonleptonic decays of the $B_s$ meson are studied 
within the framework of generalized factorization in which factorization is
applied to the tree level matrix elements while the effective Wilson
coefficients are $\mu$ and renormalization scheme independent, and
nonfactorizable effects are parametrized in terms of $\nc(LL)$ and $\nc(LR)$, 
the effective numbers of colors arising from $(V-A)(V-A)$ and $(V-A)(V+A)$
four-quark operators, respectively. Branching ratios of $B_s\to PP,PV,VV$ 
decays ($P$: pseudoscalar meson, $V$: vector meson)
are calculated as 
a function of $\nc(LR)$ with two different considerations for $\nc(LL)$:
(a) $\nc(LL)$ being fixed at the value of 2, 
and (b) $\nc(LL)=\nc(LR)$. Tree and penguin transitions are classified into
six different classes. We find that (i) the electroweak penguin
contributions account for about 85\% (for $\nc(LL)=2$) of the decay rates of
$B_s\to \eta\pi,\,\eta'
\pi,\,\eta\rho,\,\eta'\rho,\,\phi\pi,\,\phi\rho$, which receive
contributions only from tree and electroweak penguin diagrams; a measurement
of them will provide a clean determination of the electroweak penguin
coefficient $a_9$, (ii) electroweak penguin corrections to
$B_s\to\omega\etapp,\phi\eta,\omega\phi,K^{(*)}\phi,
\phi\phi$ are in general as significant as QCD penguin effects and even play
a dominant role; their 
decay rates depend strongly on $\nc(LR)$,
(iii) the branching ratio of $ B_s\to \eta\eta'$, the analogue
of $B_d\to\eta' K$, is of order $2\times 10^{-5}$, which is only slightly
larger than that of $\eta'\eta',K^{*+}\rho^-,K^+K^-,K^0\ov K^0$ decay modes,
(iv) the contribution from the $\eta'$ charm content is important
for $B_s\to\eta'\eta'$, but less significant for $B_s\to\eta\eta'$, 
and (v) the decay rates for the final states
$K^{+(*)}K^{-(*)}$ follow the pattern: $\Gamma(\ov B_s\to K^+K^-)
>\Gamma(\ov B_s\to K^+K^{*-})\gsim\Gamma(\ov B_s\to K^{*+}K^{*-})>
\Gamma(\ov B_s\to K^{+*}K^-)$ and likewise for $K^{0(*)}\ov K^{0(*)}$,
as a consequence of various interference effects between the penguin 
amplitudes governed by the effective QCD penguin coefficients $a_4$ and $a_6$.
}

\pagebreak


\section{Introduction}

Recently there has been a remarkable progress in the study of exclusive
charmless $B$ decays, both experimentally and theoretically. On the
experimental side, CLEO has discovered many new two-body decay modes 
\cite{CLEO}:
\be
 B\to\eta' K^+,~\eta' K^0_S,~\pi^\pm K^0_S,~\pi^\pm K^\mp,~\pi^0 K^\pm,~\omega
K^\pm,
\en
and a possible evidence for $B\to\phi K^*$. 
Moreover, CLEO has improved upper limits for many other channels. Therefore,
it is a field whose time has finally arrived. On the theoretical aspect, 
many important issues have been studied in past years, such as
the effective Wilson coefficients that are renormalization scale and scheme
independent, nonfactorizable effects in hadronic matrix elements, the QCD 
anomaly effect in the matrix element of pseudoscalar densities, running
light quark masses at the scale $m_b$, and the $q^2$ dependence of form 
factors.

   In the present paper, we plan to extend previous studies of charmless 
hadronic decays of $B^-,~B_d$ mesons to the $B_s$ mesons. In principle,
the physics for the $B_s$ two-body hadronic decays is very similar to 
that for the $B_d$ meson except that the spectator $d$ quark is replaced by
the $s$ quark. Experimentally, it is known that $B^\pm\to\eta' K^\pm$ and 
$B_d\to\eta' K$ have abnormally large branching ratios, several times 
larger than previous predictions. It would be very interesting to see if the
analogue of $B_d\to\eta' K$, namely $B_s\to \eta\eta'$ or $B_s\to\eta'\eta'$
still has the largest branching ratio in two-body $B_s$ charmless decays. 
Another point of interest is concerned with the electroweak penguin 
corrections. It is naively believed that in charmless $B$ decays, the
contributions from the electroweak penguin diagrams are negligible compared
to the QCD penguin corrections because of smallness of electroweak penguin
Wilson coefficients. As pointed out in \cite{Deshpande}, some
$B_s$ decay modes receive contributions only from the tree and electroweak 
penguin diagrams and moreover they are dominated by the latter. Therefore, 
electroweak penguins do play a dominant role in some of $B_s$ decays.
There also exist several penguin-dominated  $B_s$ decay modes in which 
electroweak penguin corrections to the decay rate are comparable to that 
of QCD penguin contributions. In this paper, we will study this in details.

 Experimentally, only upper limits on the branching ratios have been 
established
for a few $B_s$ rare decay modes (see \cite{PDG} or Table 7 of \cite{CLEO})
and most of them are far beyond the theoretical expectations. Nevertheless,
it is conceivable that many of the $B_s$ charmless decays can be
seen at the future hadron colliders with large $b$ production. 
Theoretically, early systematical
studies can be found in \cite{Deandrea,Xing}. More recently, one of us (B.T.)
\cite{Tseng} has analyzed the exclusive charmless $B_s$ decays involving the
$\eta$ or $\eta'$ within the framework of generalized factorization.

   This paper is organized as follows. A calculational framework is set up
in Sec.~II in which we discuss the scale and scheme independent Wilson
coefficient functions, parametrization of nonfactorizable effects, 
classification of factorizable amplitudes,..., etc. The numerical results
and discussions are presented in Sec.~III. Conclusions are summarized in
Sec.~IV. The factorizable amplitudes for all the charmless two-body $B_s$
decays are given in Appendices.

\section{Calculational framework}
\subsection{Effective Hamiltonian}
The relevant effective $\Delta B=1$ weak Hamiltonian for hadronic charmless 
$B$ decays is
\be
{\cal H}_{\rm eff}(\Delta B=1) = {G_F\over\sqrt{2}}\Big[ V_{ub}V_{uq}^*(c_1
O_1^u+c_2O_2^u)+V_{cb}V_{cq}^*(c_1O_1^c+c_2O_2^c)
-V_{tb}V_{tq}^*\sum^{10}_{i=3}c_iO_i\Big]+{\rm h.c.}, 
\en
where $q=d,s$, and
\be
&& O_1^u= (\bar ub)_\vma(\bar qu)_\vma, \qquad\qquad\qquad\qquad~~
O_2^u = (\bar qb)_\vma(\bar uu)_\vma, \non \\
&& O_{3(5)}=(\bar qb)_\vma\sum_{q'}(\bar q'q')_{\vma(\vpa)}, \qquad  \qquad
O_{4(6)}=(\bar q_\alpha b_\beta)_\vma\sum_{q'}(\bar q'_\beta q'_\alpha)_{
\vma(\vpa)},   \\
&& O_{7(9)}={3\over 2}(\bar qb)_\vma\sum_{q'}e_{q'}(\bar q'q')_{\vpa(\vma)},
  \qquad O_{8(10)}={3\over 2}(\bar q_\alpha b_\beta)_\vma\sum_{q'}e_{q'}(\bar 
q'_\beta q'_\alpha)_{\vpa(\vma)},   \non
\en   
with $O_3$--$O_6$ being the QCD penguin operators, $O_{7}$--$O_{10}$ 
the electroweak penguin operators, and $(\bar q_1 q_2)_{_{V\pm A}}\equiv
\bar q_1\gamma_\mu(1\pm \gamma_5)q_2$.
In order to ensure the renormalization-scale and -scheme
independence for the physical amplitude, the matrix element of 4-quark
operators has to be evaluated in the same renormalization scheme as
that for Wilson coefficients $c_i(\mu)$ and renormalized at the same 
scale $\mu$. Generically, the hadronic matrix element is related to the tree
level one via
\be
\la O(\mu)\ra=g(\mu)\la O\ra_{\rm tree},
\en
with $g(\mu)$ being the perturbative corrections to the four-quark operators
renormalized at the scale $\mu$. We employ the relation (2.3) to write
$\la{\cal H}_{\rm eff}\ra
=c^{\rm eff}\la O\ra_{\rm tree}$. Schematically, the effective Wilson
coefficients are given by $c^{\rm eff}=c(\mu)g(\mu)$. Formally, one can
show that $c_i^{\rm eff}$ are $\mu$ and renormalization scheme independent. It 
is at this stage that the factorization approximation is applied to 
the hadronic matrix elements of the operator $O$ at the tree level. The
physical amplitude obtained in this manner is guaranteed to be renormalization
scheme and scale independent.
\footnote{This formulation is different from the one advocated in 
\cite{Neubert} in which the $\mu$ dependence of the Wilson coefficients
$c_i(\mu)$ are assumed to be canceled out by that of the nonfactorization 
parameters $\vp_8(\mu)$ and $\vp_1(\mu)$ so that the effective 
parameters $a_i^{\rm eff}$ are $\mu$ independent.}

Perturbative QCD and electroweak corrections to $g(\mu)$ 
from vertex diagrams and penguin diagrams have been
calculated in \cite{Buras92,Flei,Kramer,Ali}.
The penguin-type corrections  depend on $k^2$, the
gluon's momentum squared, so are the effective Wilson coefficient functions. 
To the next-to-leading order, we obtain \cite{CT98}
\be
&& {c}_1^{\rm eff}=1.149, \qquad\qquad\qquad\qquad\qquad {c}_2^{\rm eff}=
-0.325,   \non \\
&& {c}_3^{\rm eff}=0.0211+i0.0045, \qquad\qquad\quad\,{c}_4^{\rm eff}=-0.0450
-i0.0136, \non  \\
&& {c}_5^{\rm eff}=0.0134+i0.0045, \qquad\qquad\quad\,{c}_6^{\rm eff}=-0.0560
-i0.0136, \non  \\
&& {c}_7^{\rm eff}=-(0.0276+i0.0369)\alpha, \qquad \quad {c}_8^{\rm eff}=
0.054\,\alpha, \non  \\
&& {c}_9^{\rm eff}=-(1.318+i0.0369)\alpha, \qquad \quad ~\,{c}_{10}^{\rm eff}=
0.263\,\alpha,
\en
at $k^2=m^2_b/2$. It is interesting to note that $c_{1,2}^{\rm eff}$ 
are very close to the leading order Wilson coefficients: 
$c_1^{\rm LO}=1.144$ and $c_2^{\rm LO}=-0.308$ at $\mu={m}_b(m_b)$ 
\cite{Buras96} 
and that Re$(c^{\rm eff}_{3-6})\approx {3\over 2} c_{3-6}^{\rm LO}(\mu)$. 
Therefore, the decay rates of charmless $B$ decay modes
dominated by QCD penguin diagrams will be too small by a factor of 
$\sim (1.5)^2=2.3$ if 
only leading-order penguin coefficients are employed for the calculation.

\subsection{Parametrization of nonfactorizable effects}
Because there is only one single form factor (or Lorentz scalar) 
involved in the class-I or class-II decay amplitude of $B\to PP,~PV$ decays
(see Sec.~II.C for the classification of factorizable amplitudes), 
the effects of nonfactorization can be lumped into the
effective parameters $a_1$ and $a_2$ \cite{Cheng94}:
\be
a_1^{\rm eff}=c_1^{\rm eff}+c_2^{\rm eff}\left({1\over N_c}+\chi_1\right),
\qquad a_2^{\rm eff}=c_2^{\rm eff}+c_1^{\rm eff}\left({1\over N_c}
+\chi_2\right),
\en
where $\chi_i$ are nonfactorizable terms and receive main contributions from
color-octet current operators. 
Since $|c^{\rm eff}_1/c^{\rm eff}_2|\gg 1$, it is evident from Eq.~(2.5) 
that even a small amount of nonfactorizable contributions will have a
significant effect on the color-suppressed class-II amplitude.
If $\chi_{1,2}$ are universal (i.e. process independent) in
charm or bottom decays, then we have a generalized factorization scheme
in which the decay amplitude is expressed in terms of factorizable 
contributions multiplied by the universal effective parameters
$a_{1,2}^{\rm eff}$. For $B\to VV$ decays, this new factorization implies
that nonfactorizable terms contribute in equal weight to all partial wave
amplitudes so that $a_{1,2}^{\rm eff}$ {\it can} be defined.
It should be stressed that, contrary to the
naive one, the improved factorization does incorporate nonfactorizable
effects in a process independent form. For example, $\chi_1=\chi_2=-{1
\over 3}$ in the large-$N_c$ approximation of factorization.
Phenomenological analyses
of the two-body decay data of $D$ and $B$ mesons indicate that while
the generalized factorization hypothesis in general works reasonably well, 
the effective parameters $a_{1,2}^{\rm eff}$ do show some variation from
channel to channel, especially for the weak decays of charmed mesons
 \cite{Cheng94,Kamal96,Cheng96}.
An eminent feature emerged from the data analysis is that $a_2^{\rm eff}$ 
is negative in charm decay, whereas it becomes positive in the two-body decays
of the $B$ meson \cite{Cheng94,CT95,Neubert}:
\be
a_2^{\rm eff}(D\to\overline{K}\pi)\sim -0.50\,, \qquad a_2^{\rm eff}(B\to 
D\pi)\sim 0.20-0.28\,.
\en
It should be stressed that the magnitude of $a_{1,2}$ depends on the 
model results for form factors. It follows that
\be
\chi_2(D\to \overline{K}\pi)\sim -0.36\,, \qquad \chi_2(B\to D\pi)\sim 
0.12-0.19\,.
\en
The observation $|\chi_2(B)|\ll|\chi_2(D)|$ is consistent
with the intuitive picture that soft gluon effects become stronger when the
final-state particles move slower, allowing more time for significant
final-state interactions after hadronization \cite{Cheng94}.
Phenomenologically, it is often to treat the number of colors $N_c$ as
a free parameter to model the nonfactorizable contribution to hadronic
matrix elements and its value can be extracted from the data of two-body
nonleptonic decays. Theoretically, this amounts to
defining an effective number of colors $\nc$, called $1/\xi$ in \cite{BSW87},
by 
\be
1/N_c^{\rm eff}\equiv (1/N_c)+\chi\,.
\en
It is clear from (2.7) that
\be
N_c^{\rm eff}(D\to \overline{K}\pi) \gg 3,\qquad N_c^{\rm eff}(B\to D\pi)
\sim 1.8-2.2\,.
\en

  The effective Wilson coefficients appear in the factorizable decay 
amplitudes in the combinations $a_{2i}=
{c}_{2i}^{\rm eff}+{1\over N_c}{c}_{2i-1}^{\rm eff}$ and $a_{2i-1}=
{c}_{2i-1}^{\rm eff}+{1\over N_c}{c}^{\rm eff}_{2i}$ $(i=1,\cdots,5)$. 
As discussed in the Introduction, 
nonfactorizable effects in the 
decay amplitudes of $B\to PP,~VP$ can be absorbed into the parameters
$a_i^{\rm eff}$. This amounts to replacing $N_c$ in $a^{\rm eff}_i$
by $(N_c^{\rm eff})_i$. Explicitly,
\be
a_{2i}^{\rm eff}={c}_{2i}^{\rm eff}+{1\over (N_c^{\rm eff})_{2i}}{c}_{2i-1}^{
\rm eff}, \qquad \quad a_{2i-1}^{\rm eff}=
{c}_{2i-1}^{\rm eff}+{1\over (N_c^{\rm eff})_{2i-1}}{c}^{\rm eff}_{2i}, \qquad
(i=1,\cdots,5).
\en
It is customary to assume in the literature that $(N_c^{\rm eff})_1
\approx (N_c^{\rm eff})_2\cdots\approx (N_c^{\rm eff})_{10}$
so that the subscript $i$ can be dropped; that is, the nonfactorizable
term is usually assumed to behave in the same way in penguin and tree
decay amplitudes. A closer investigation shows 
that this is not the case. We have argued in \cite{CT98} that
nonfactorizable effects in the matrix
elements of $(V-A)(V+A)$ operators are {\it a priori} different from that of 
$(V-A)(V-A)$ operators. One reason is that
the Fierz transformation of the $(V-A)(V+A)$ operators $O_{5,6,7,8}$
is quite different from that of $(V-A)(V-A)$ operators $O_{1,2,3,4}$
and $O_{9,10}$. As a result, contrary to the common assumption,
$\nc(LR)$ induced by the $(V-A)(V+A)$ operators are
theoretically different from $\nc(LL)$ generated by
the $(V-A)(V-A)$ operators \cite{CT98}. From Eq.~(2.10) it is expected that
\be
&& N_c^{\rm eff}(LL)\equiv
\left(N_c^{\rm eff}\right)_1\approx\left(N_c^{\rm eff}\right)_2\approx
\left(N_c^{\rm eff}\right)_3\approx\left(N_c^{\rm eff}\right)_4\approx
\left(N_c^{\rm eff}\right)_9\approx
\left(N_c^{\rm eff}\right)_{10},   \non\\
&& N_c^{\rm eff}(LR)\equiv
\left(N_c^{\rm eff}\right)_5\approx\left(N_c^{\rm eff}\right)_6\approx
\left(N_c^{\rm eff}\right)_7\approx
\left(N_c^{\rm eff}\right)_8,
\en
and $N_c^{\rm eff}(LR)\neq N_c^{\rm eff}(LL)$ in general. In principle, 
$N_c^{\rm eff}$ can vary from channel to channel, as in the case of charm
decay. However, in the energetic two-body $B$ decays, $\nc$
is expected to be process insensitive as supported by data 
\cite{Neubert}. 

   The $\nc$-dependence of the effective parameters $a_i^{\rm eff}$'s
are shown in Table I for several representative values of $\nc$. From 
Table I we see that (i) the dominant coefficients are $a_1,\,a_2$ for
current-current amplitudes, $a_4$ and $a_6$ for QCD penguin-induced
amplitudes, and $a_9$ for electroweak penguin-induced amplitudes,
and (ii) $a_1,a_4,a_6$ and $a_9$ are $\nc$-stable, while others depend
strongly on $\nc$. Therefore, for charmless $B$ decays whose decay
amplitudes depend dominantly on $\nc$-stable coefficients, their decay
rates can be reliably predicted within the factorization approach even in
the absence of information on nonfactorizable effects.

  The CLEO data of $B^\pm\to\omega\pi^\pm$ 
available last year clearly indicate that $\nc(LL)$ is favored to be small, 
$\nc(LL)<2.9\,$ \cite{CT98}. If the value of $\nc(LL)$ is fixed to be 2, 
the branching ratio of $B^\pm\to\omega\pi^\pm$ for positive $\rho$ ($\rho$
being a Wolfenstein parameter; see Sec.~II.D),
which is preferred by the current 
analysis \cite{Parodi}, will be of order $(0.9-1.0)\times 10^{-5}$, which is 
very close to the central value of the measured one.
Unfortunately, the significance 
of $B^\pm\to\omega\pi^\pm$ is reduced in the recent CLEO analysis and only an
upper limit is quoted \cite{CLEOomega2}.
Nevertheless, the central value of ${\cal B}(B^\pm\to
\pi^\pm\omega)$ remains about the same.
Therefore, a measurement of its branching ratio is
urgently needed. A very recent CLEO analysis of $B^0\to\pi^+\pi^-$ \cite{Roy}
presents an improved upper limit, ${\cal B}(B^0\to\pi^+\pi^-)<0.84\times 
10^{-5}$. 
If the form factor $F_0^{B\pi}(0)$ is known, this tree-dominated decay 
could offer a useful constraint on $\nc(LL)$ as its branching ratio increases 
slightly with $\nc$. For $F_0^{B\pi}(0)=0.30$, we find $\nc(LL)\lsim 2.0$.
The fact that $\nc(LL)$ is favored to be at the value of 2
in hadronic charmless two-body decays of the $B$ meson 
is consistent with the nonfactorizable term extracted from $B\to (D,D^*) 
\pi,~D\rho$ decays, namely $\nc(B\to D\pi)\approx 2$. 
Since the energy release in the energetic two-body decays $B\to\omega\pi$, 
$B\to D\pi$
is of the same order of magnitude, it is thus expected that $\nc(LL)|_{B
\to\omega\pi}\approx 2$.
In analogue to the class-III $B\to D\pi$ decays, the
interference effect of spectator amplitudes in charged $B$ decays 
$B^-\to\pi^-\pi^0,~\rho^-\pi^0,~\pi^-\rho^0$ is sensitive to $\nc(LL)$;
measurements of them will be very useful to pin down 
the value of $\nc(LL)$.

As for $\nc(LR)$, it is found in \cite{CT98} that the constraints on $\nc(LR)$
derived from $B^\pm\to\phi K^\pm$ and $B\to\phi K^*$ are not consistent. 
Under the factorization hypothesis, the decays $B\to\phi K$ and 
$B\to\phi K^*$ should have almost the same branching ratios, a prediction not
borne out by current data. Therefore, it is crucial to measure the charged
and neutral decay modes of $B\to\phi(K,K^*)$ in order to see if
the generalized factorization approach is applicable to $B\to\phi K^*$ decay.
Nevertheless, the analysis of $B\to\eta' K$ in \cite{CT98} indicates that 
$\nc(LL)\approx 2$
is favored and $\nc(LR)$ is preferred to be larger. Since the energy 
release in the energetic two-body charmless $B$ decays is not less
than that in $B\to D\pi$ decays, it is thus expected that
\be
|\chi({\rm 2-body~rare~B~decay})|\lsim |\chi(B\to D\pi)|.
\en
It follows from Eqs.~(2.7) and (2.8) that $\nc(LL)\approx \nc(B\to D\pi)\sim 
2$ and $\nc(LR)\sim 2-5$, depending on the sign of $\chi$. Since $\nc(LR)>
\nc(LL)$ implied by the data, therefore, we conjecture that 
\be
\nc(LL)\approx 2, \qquad \quad \nc(LR)\lsim 5.
\en

\begin{table}[ht]
{\small Table I. Numerical values for the effective coefficients 
$a_i^{\rm eff}$ at $\nc=2,3,5,\infty$ (in units of $10^{-4}$ for
$a_3,\cdots,a_{10}$). For simplicity we will drop the superscript ``eff'' 
henceforth.}
\begin{center}
\begin{tabular}{ c c c c c }
 & $N_c^{\rm eff}=2$ & $N_c^{\rm eff}=3$ & $N_c^{\rm eff}=5$  & 
$N_c^{\rm eff}=\infty$ \\ \hline
$a_1$  &0.986 & 1.04  & 1.08 &  1.15  \\
$a_2$  &0.25 &  0.058 & --0.095 &  --0.325 \\
$a_3$  &$-13.9-22.6i$   & 61 & $121+18.1i$ &$211+45.3i$ \\
$a_4$  &$-344-113i$ & $-380-121i$  & $-408-127i$ & $-450-136i$ \\
$a_5$  &$-146-22.6i$ & $-52.7$  & $22.0+18.1i$ & $134+ 45.3i$ \\
$a_6$  &$- 493-113i$ & $-515-121i$  & $-533-127i$ & $-560-136i$ \\
$a_7$  &$ 0.04-2.73i$ & $-0.71-2.73i$  & $-1.24-2.73i$ & $-2.04-2.73i$  \\
$a_8$  &$2.98 -1.37i$ & $3.32-0.91i$  & $3.59-0.55i$ & 4   \\
$a_9$  &$-87.9- 2.73i$ & $-91.1-2.73i$  & $-93.7-2.73i$ & $-97.6-2.73i$ \\
$a_{10}$&$-29.3-1.37i$ & $-13.1-0.91i$  & $-0.04-0.55i$ & 19.48   \\
\end{tabular}
\end{center}
\end{table}

\subsection{Factorizable amplitudes and their classification}
Applying the effective Hamiltonian (2.1), the factorizable decay amplitudes 
of $\ov B_s\to PP,VP,VV$ obtained  within the generalized factorization 
approach are summarized in the Appendices A,B,C, where, for simplicity, we 
have neglected $W$-annihilation, space-like penguins and final-state
interactions. All the penguin contributions
to the decay amplitudes can be derived from Table II by studying the
underlying $b$ quark weak transitions. To illustrate this, 
let $X^{(BM_1,M_2)}$
denote the factorizable amplitude with the meson $M_2$ being factored out:
\be
X^{(BM_1,M_2)}=\la M_2|(\bar q_2q_3)\vma|0\ra\la M_1|(\bar q_1b)_\vma|\ov
B\ra.
\en
In general, when $M_2$ is a charged state, only $a_{\rm even}$ penguin terms
contribute. For example, from Table II we obtain
\be
A(\ov B_s\to K^+\pi^-)_{\rm peng} &\propto& \left[ a_4+a_{10}+(a_6+a_8)R
\right]X^{(B_s K^+,\pi^-)},    \non \\
A(\ov B_s\to K^{*+}\pi^-)_{\rm peng} &\propto& \left[ a_4+a_{10}-(a_6+a_8)R'
\right]X^{(B_s K^{*+},\pi^-)},    \non \\
A(\ov B_s\to K^+\rho^-)_{\rm peng} &\propto& \left[ a_4+a_{10}
\right]X^{(B_s K^+,\rho^-)},    
\en
with $R'\approx R\approx m_\pi^2/(m_bm_d)$. 
When $M_2$ is a neutral meson with $I_3=0$, namely, $M_2=\pi^0,\rho^0,\omega$ 
and $\etapp$, $a_{\rm odd}$ penguin terms start to contribute.  
From Table II we see that the decay amplitudes of 
$\ov B_s\to M\pi^0,~\ov B_s
\to M\rho^0,~\ov B_s\to M\omega,~\ov B_s\to M\etapp$ contain the following
respective factorizable terms:
\be
&& {3\over 2}(-a_7+a_9)X_u^{(B_sM,\pi^0)},   \non \\
&& {3\over 2}(a_7+a_9)X_u^{(B_sM,\rho^0)},   \non \\
&& (2a_3+2a_5+{1\over 2}a_7+{1\over 2}a_9)X_u^{(B_sM,\omega)},   \non \\
&& (2a_3-2a_5-{1\over 2}a_7+{1\over 2}a_9)X_u^{(B_sM,\etapp)},   
\en
where the subscript $u$ indicates the $u\bar u$ quark content of the neutral
meson:
\be
X^{(B_sM,\pi^0)}_u=\la \pi^0|(\bar uu)\vma|0\ra\la M_1|(\bar q_1b)_\vma|\ov
B_s\ra.
\en
For example, the penguin amplitudes of $\ov B_s\to \eta\omega $ and $K^0\pi^0$
are given by
\be
A(\ov B_s\to \eta\omega)_{\rm peng} &\propto & \left[2(a_3+a_5)+{1\over 2}
(a_7+a_9)\right]X_u^{(B_s\eta,\omega)},   \non \\
A(\ov B_s\to K^0\pi^0)_{\rm peng} &\propto &  {3\over 2}(-a_7+a_9)X_u^{
(B_s K^0,
\pi^0)} +\Big[a_4-{1\over 2}a_{10}+(a_6-{1\over 2}a_8)R\Big]X_d^{(B_s K^0,
\pi^0)},  \non \\
&\propto& \left[ -a_4+{3\over 2}(-a_7+a_9)+{1\over 2}a_{10}-(a_6-{1\over 2}
a_8)R\right]X_u^{(B_sK^0,\pi^0)},
\en
respectively. It is interesting to note that
the decays $\ov B_s\to (\etapp,\phi)(\pi^0,\rho^0)$ do not receive 
any contributions from QCD penguin diagrams and they are dominated by 
electroweak penguins. We will come back to this interesting observation later.

\vskip 0.4cm
\begin{table}[ht]
{\small Table II. Penguin contributions to the factorizable $B\to PP,~VP,VV$ 
decay amplitudes multiplied by $-(G_F/\sqrt{2})V_{tb}V_{tq}^*$, where
$q=d,s$. The notation $B\to M_1,M_2$ means that the meson $M_2$ can be factored
out under the factorizable approximation. In addition to the $a_{\rm even}$ 
terms, the decay also receives contributions from $a_{\rm odd}$ penguin
effects when $M_2$ is 
a neutral meson with $I_3=0$. Except for $\eta$ or
$\eta'$ production, the coefficients $R$ and $R'$ are given by
$R=2m_P^2/[(m_1+m_2)(m_b-m_3)]$ and $R'=-2m_P^2/[(m_1+m_2)(m_b+m_3)]$,
respectively.}
\begin{center}
\begin{tabular}{ c c c }
Decay  & $b\to qu\bar u,~b\to qc\bar c$ & $b\to qd\bar d,~b\to qs\bar s$  
\\ \hline
$B\to P,P$ & $a_4+a_{10}+(a_6+a_8)R$ & $a_4-{1\over 2}a_{10}+(a_6-{1\over 2}
a_8)R$ \\
$B\to V,P$ & $a_4+a_{10}+(a_6+a_8)R'$ & $a_4-{1\over 2}a_{10}+(a_6-{1\over 2}
a_8)R'$ \\
$B\to P,V$ & $a_4+a_{10}$ & $a_4-{1\over 2}a_{10}$ \\
$B\to V,V$ & $a_4+a_{10}$ & $a_4-{1\over 2}a_{10}$ \\  \hline
$B\to P,P^0$ & $a_3-a_5-a_7+a_9$ & $a_3-a_5+{1\over 2}a_7-{1\over 2}a_9$ \\
$B\to V,P^0$ & $a_3-a_5-a_7+a_9$ & $a_3-a_5+{1\over 2}a_7-{1\over 2}a_9$ \\
$B\to P,V^0$ & $a_3+a_5+a_7+a_9$ & $a_3+a_5-{1\over 2}a_7-{1\over 2}a_9$ \\
$B\to V,V^0$ & $a_3+a_5+a_7+a_9$ & $a_3+a_5-{1\over 2}a_7-{1\over 2}a_9$ \\
\end{tabular}
\end{center}
\end{table}

   Just as the charm decays or $B$ decays into the charmed meson, the 
tree-dominated amplitudes for hadronic charmless $B$ decays are customarily 
classified into three classes \cite{BSW87}:
\begin{itemize}
\item Class-I for the decay modes dominated by the external $W$-emission
characterized by the parameter $a_1$. Examples are $\ov B_s\to 
K^+\pi^-,\,K^{*+}\pi^-,\cdots$.
\item Class-II for the decay modes dominated by the color-suppressed 
internal $W$-emission characterized by the parameter $a_2$. Examples are
$\ov B_s\to K^0\pi^0,\,K^0\rho^0,\cdots$.
\item Class-III decays involving both external and internal $W$ emissions.
Hence the class-III amplitude is of the form $a_1+ra_2$. This class does
not exist for the $B_s$.
\end{itemize}
Likewise, penguin-dominated charmless $B_s$ decays can be classified 
into three categories:
\footnote{Our classification of factorizable penguin amplitudes is not the 
same as that in \cite{Ali3}; we introduce three new classes in the same spirit
as the classification of tree-dominated decays.}
\begin{itemize}
\item Class-IV for those decays whose amplitudes are governed by the QCD 
penguin parameters $a_4$ and $a_6$ in the combination $a_4+Ra_6$, where the
coefficient $R$ arises from the $(S-P)(S+P)$ part of the operator $O_6$.
In general, $R=2m_{P_b}^2/[(m_1+m_2)(m_b-m_3)]$ for $B\to P_aP_b$ with the
meson $P_b$ being factored out under the factorizable approximation,
$R=-2m_{P_b}^2/[(m_1+m_2)(m_b+m_3)]$ for $B\to V_aP_b$, and $R=0$ for
$B\to P_aV_b$ and $B\to V_aV_b$. Note that $a_4$ is always accompanied
by $a_{10}$, and $a_6$ by $a_8$. In short, class-IV modes are governed by
$a_{\rm even}$ penguin terms. Examples are $\ov B_s\to K^+K^-,\,K^0
\ov K^0,\,\phi\etapp,\cdots$.
\item Class-V modes for those decays whose amplitudes are governed by the
effective coefficients $a_3,a_5,a_7$ and $a_9$ (i.e. $a_{\rm odd}$ penguin
terms) in the combinations $a_3\pm a_5$ and/or $a_7\pm a_9$ (see Table II).  
Examples are $\ov B_s\to \pi\etapp,\,\omega\etapp,\,\pi\phi,\cdots$.
\item Class-VI involving the interference of class-IV and class-V decays,
e.g. $\ov B_s\to \etapp\etapp,\phi\etapp,K^0\phi,\cdots$.
\end{itemize}

    Sometimes the tree and penguin contributions are comparable. In 
this case, the interference between penguin and spectator amplitudes
is at work. There are three such decays: $\ov B_s\to K^0\omega,K^{*0}\etapp,
K^{*0}\omega$; they involve class-II and -VI amplitudes (see Tables IV and V).

\subsection{Input parameters}
In this subsection we specify the values for various parameters employed in 
the present paper. For current quark masses, we employ the running masses at 
the scale $\mu=m_b$:
\be
&& m_u(m_b)=3.2\,{\rm MeV},  \qquad m_d(m_b)=6.4\,{\rm MeV},  \qquad
m_s(m_b)=105\,{\rm MeV},  \non \\
&&  m_c(m_b)=0.95\,{\rm GeV},  \qquad
m_b(m_b)=4.34\,{\rm GeV}.
\en
As for the Wolfenstein parameters $A,\lambda,\rho$ and $\eta$, which are 
utilized to parametrize the quark mixing matrix, we use $A=0.804,~\lambda=
0.22,~\rho=0.16$ and $\eta=0.34$. The values for $\rho$ and $\eta$ follow from
a recent analysis of all available experimental 
constraints imposed on the Wolfenstein parameters \cite{Parodi}:
\be
\bar{\rho}=\,0.156\pm 0.090\,,   \qquad \bar{\eta}=\,0.328\pm 0.054,
\en
where $\bar{\rho}=\rho(1-{\lambda^2\over 2})$ and $\bar\eta=\eta(1-{\lambda
^2\over 2})$. For the values of decay constants, we use $f_\pi=132$ MeV, 
$f_K=160$ MeV, $f_\rho=210$ MeV, $f_{K^*}=221$ MeV, $f_\omega=195$ MeV and
$f_\phi=237$ MeV.

   To determine the decay constant $f_{\etapp}^q$, defined by
$\la 0|\bar q\gamma_\mu\gamma_5 q|\etapp\ra=if_{\etapp}^q p_\mu$, it has
been emphasized \cite{Leutwyler,Kroll2}
that the decay constants do not simply follow the $\eta-\eta'$
state mixing given by 
\be
\eta'=\eta_8\sin\theta+\eta_0\cos\theta, \qquad \eta=\eta_8\cos\theta-\eta_0
\sin\theta.
\en
Introduce the decay constants $f_8$ and $f_0$ by 
\be
\la 0|A_\mu^0|\eta_0\ra=if_0 p_\mu, \qquad \la 0|A_\mu^8|\eta_8\ra=if_8 p_\mu.
\en
Because of SU(3) breaking, the matrix elements $\la 0|A_\mu^{0(8)}|\eta_{8(0)}
\ra$ do not vanish in general and they will induce a two-angle mixing 
among the decay constants, that is,  $f_{\eta'}^u$ and $f_{\eta'}^s$ are 
related to $f_8$ and $f_0$ by
\be
f_{\eta'}^u={f_8\over\sqrt{6}}\sin\theta_8+{f_0\over\sqrt{3}}\cos\theta_0,
\qquad f_{\eta'}^s=-2{f_8\over\sqrt{6}}\sin\theta_8+{f_0\over\sqrt{3}}\cos
\theta_0.     
\en
Likewise,
\be
f_{\eta}^u={f_8\over\sqrt{6}}\cos\theta_8-{f_0\over\sqrt{3}}\sin\theta_0,
\qquad f_{\eta}^s=-2{f_8\over\sqrt{6}}\cos\theta_8-{f_0\over\sqrt{3}}\sin
\theta_0.
\en
Based on the ansatz that the decay constants in the quark flavor basis
follow the pattern of particle state mixing,
relations between $\theta_8,~\theta_0$ and $\theta$
are derived in \cite{Kroll2}, where $\theta$ is the $\eta-\eta'$ mixing angle
introduced in (2.21). It is found in \cite{Kroll2} that
phenomenologically 
\be
\theta_8=-21.2^\circ, \qquad \theta_0=-9.2^\circ, \qquad \theta=-15.4^\circ,
\non \\
\en
and
\be
f_8/f_\pi=1.26,  \qquad f_0/f_\pi=1.17.
\en
The decay constant $f_{\eta'}^c$, 
defined by $\la 0|\bar c\gamma_\mu\gamma_5c|\eta'\ra=if_{\eta'}^c
q_\mu$, has been determined from theoretical calculations 
\cite{Halperin,Ali2,Araki} and from the
phenomenological analysis of the data of $J/\psi\to\eta_c\gamma,\,J/\psi
\to\eta'\gamma$ and of the $\eta\gamma$ and $\eta'\gamma$ transition form
factors \cite{Ali,Kroll2,Petrov,Kroll,Cao}; it lies in the range
--2.3 MeV $\leq f_{\eta'}^c\leq$ --18.4 MeV. In this paper we use the values
\be
f_{\eta'}^c=-(6.3\pm 0.6)\,{\rm MeV},  \qquad f_\eta^c=-(2.4\pm 0.2)\,{\rm 
MeV},
\en
as obtained in \cite{Kroll2}.

  For form factors, the Bauer-Stech-Wirbel (BSW) model \cite{BSW85} gives 
\cite{Xing}
\footnote{The form factors adopted in \cite{Tseng} are calculated using
the light-front quark model and in general they are larger than the BSW 
model's results.}
\be
F_0^{B_sK}(0)=0.274, \quad && F_0^{B_s\eta_{s\bar s}}(0)=0.335, \qquad
F_0^{B_s\eta'_{s\bar s}}(0)=0.282,  \non \\
A_0^{B_s\phi}(0)=0.272, \quad && ~~A_1^{B_s\phi}(0)=0.273, \qquad
A_2^{B_s\phi}(0)=0.273, \non \\
A_0^{B_s K^*}(0)=0.236, \quad && A_1^{B_s K^*}(0)=0.232, \qquad
A_2^{B_s K^*}(0)=0.231, \non \\
V^{B_s\phi}(0)=0.319, \quad && V^{B_s K^*}(0)=0.281.
\en
It should be stressed that the $\eta-\eta'$ wave function 
normalization has not been included in the form factors 
$F_0^{B_s\eta_{s\bar s}}$ and $F_0^{B_s\eta'_{s\bar s}}$; they are 
calculated in a relativistic quark model by putting
the $s\bar s$ constitutent quark mass only.
To compute the physical 
form factors, one has to take into account the wave function
normalizations of the $\eta$ and $\eta'$:
\be
F_0^{B_s\eta}=-\left({2\over\sqrt{6}}\cos\theta+{1\over\sqrt{3}}\sin\theta
\right)F_0^{B_s\eta_{s\bar s}},  \qquad &&
F_0^{B_s\eta'}=\left(-{2\over\sqrt{6}}\sin\theta+{1\over\sqrt{3}}\cos\theta
\right)F_0^{B_s\eta'_{s\bar s}}.
\en
It is clear that the form factors $F_0^{B_s\eta}$ and $F_0^{B_s\eta'}$ have
opposite signs.

For the $q^2$ dependence of form factors in the region where $q^2$ is not 
too large, we shall use the pole dominance ansatz, namely,
\be
f(q^2)=\,{f(0)\over \left(1-{q^2/m^2_*}\right)^n},
\en
where $m_*$ is the pole mass given in \cite{BSW87}. 
A direct calculation of $B\to P$ and $B\to V$ 
form factors at time-like momentum transfers is available in the relativistic 
light-front quark model \cite{CCH} with the results that
the $q^2$ dependence of the form factors $A_0,~A_2,~V,~~F_1$ is a dipole 
behavior (i.e. $n=2$), while $F_0,~A_1$ exhibit a monopole dependence ($n=1$).

Recently, the $B_s\to K^*$ and $B_s\to \phi$ form factors have also been
calculated in the light-cone sum rule approach \cite{Ball} with the 
parametrization
\be
f(q^2)=\,{f(0)\over 1-a(q^2/m_{B_s}^2)+b(q^2/m_{B_s}^2)^2}
\en
for the form-factor $q^2$ dependence. The results are \cite{Ball}
\be
A_0^{B_s\phi}(0)=0.382, \qquad && a=1.77, \qquad b=0.856,   \non \\
A_1^{B_s\phi}(0)=0.296, \qquad && a=0.87, \qquad b=-0.061,   \non \\
A_2^{B_s\phi}(0)=0.255, \qquad && a=1.55, \qquad b=0.513,   \non \\
V^{B_s\phi}(0)=0.433, \qquad && a=1.75, \qquad b=0.736,  \non \\
A_0^{B_s K^*}(0)=0.254, \qquad && a=1.87, \qquad b=0.887,   \non \\
A_1^{B_s K^*}(0)=0.190, \qquad && a=1.02, \qquad b=-0.037,   \non \\
A_2^{B_s K^*}(0)=0.164, \qquad && a=1.77, \qquad b=0.729,   \non \\
V^{B_s K^*}(0)=0.262, \qquad && a=1.89, \qquad b=0.846\,.  
\en
It is obvious that the $q^2$ dependence for the form factors $A_0,A_2$ and $V$ 
is dominated by the dipole terms, while $A_1$  by the monopole term in the
region where $q^2$ is not too 
large. In Tables IV and V we will present results using these two different
parametrizations for $B_s\to V$ form factors.

  We will encounter matrix elements of pseudoscalar densities when evaluating
the penguin amplitudes. Care must be taken to consider the pseudoscalar
matrix element for $\etapp\to$ vacuum transition: The anomaly effects must be
included in order to ensure a correct chiral behavior for the pseudoscalar
matrix element \cite{CT98}. The results are \cite{Kagan,Ali}
\be
\la\etapp|\bar s\gamma_5 s|0\ra &=& -i{m_{\etapp}^2\over 2m_s}\,\left(f_{
\etapp}^s-f^u_{\etapp}\right),  \non \\
\la\etapp|\bar u\gamma_5u|0\ra &=& \la\etapp|\bar d\gamma_5d|0\ra=r_{\etapp}
\,\la\etapp|\bar s\gamma_5s|0\ra,
\en
with \cite{CT98}
\be
r_{\eta'} &=& {\sqrt{2f_0^2-f_8^2}\over\sqrt{2f_8^2-f_0^2}}\,{\cos\theta+
{1\over \sqrt{2}}\sin\theta\over \cos\theta-\sqrt{2}\sin\theta}, \non \\
r_{\eta} &=& -{1\over 2}\,{\sqrt{2f_0^2-f_8^2}\over\sqrt{2f_8^2-f_0^2}}\,
{\cos\theta-\sqrt{2}\sin\theta\over \cos\theta+{1\over\sqrt{2}}
\sin\theta}.
\en

\section{Numerical results and discussions}
   With the factorizable decay amplitudes summarized in Appendices and the 
input parameters shown in Sec.~II, we are ready to compute the branching ratios
for the
two-body charmless nonleptonic decays of the $B_s$ meson. The decay rates
for $B_s\to PP,VP$ are given by
\be
\Gamma(B_s\to P_1P_2) &=& {p_c\over 8\pi m^2_{B_s}} |A(B_s\to P_1P_2)|^2,   
\non\\
\Gamma(B_s\to VP) &=& {p_c^3\over 8\pi m^2_V} |A(B_s\to VP)/(\vp\cdot 
p_{_{B_s}})|^2.
\en
The decay $B_s\to VV$ is more complicated as its amplitude involves three 
form factors. In general, the factorizable amplitude of $B_s\to V_1V_2$ is
of the form:
\be
A(B_s\to V_1V_2) &=& \alpha X^{(B_sV_1,V_2)}+\beta X^{(B_sV_2,V_1)}   \non \\
&=& (\alpha_1 A_1^{B_sV_1}+\beta_1A_1^{B_sV_2})\vp^*_1\cdot\vp_2^*
+(\alpha_2 A_2^{B_sV_1}+\beta_2 A_2^{B_sV_2})(\vp^*_1\cdot p_{_{B_s}})(\vp_2
^*\cdot p_{_{B_s}})   \non \\
&+& i\vp_{\mu\nu\rho\sigma}\vp^{*\mu}_2\vp^{*\nu}_1p^\rho_{_{B_s}}p^\sigma_1
(\alpha_3 V^{B_sV_1}+\beta_3 V^{B_sV_2}),   
\en
where use of Eq.~(C1) has been made. Then 
\be
\Gamma(B_s\to V_1V_2)={p_c\over 8\pi m^2_{_{B_s}} }|\alpha_1 (m_{B_s}+m_1)
m_2f_{V_2}A_1^{B_sV_1}(m^2_2)|^2(H+2\zeta H_1+2\zeta^2 H_2),
\en
where
\be
H &=& (a-bx)^2+2(1+c^2y^2),   \non \\
H_1 &=& (a-bx)(a-b'x')+2(1+cc'yy'),   \non \\
H_2 &=& (a-b'x')^2+2(1+c'^2y'^2),
\en
with
\be
&&  a={m_{B_s}^2-m_1^2-m_2^2\over 2m_1m_2}, \qquad b={2m^2_{B_s}p_c^2\over
m_1m_2(m_{B_s}+m_1)^2}, \qquad c={2m_{B_s}p_c\over (m_{B_s}+m_1)^2}, \non\\
&& \zeta={\beta_1A_1^{B_s V_2}(m_1^2)\over \alpha_1 A_1^{B_sV_1}(m_2^2)},
\qquad \quad x={A_2^{B_sV_1}(m_2^2)\over A_1^{B_sV_1}(m^2_2)}, \qquad \quad
y={V^{B_sV_1}(m_2^2)\over A_1^{B_sV_1}(m_2^2)},
\en 
where $p_c$ is the c.m. momentum, $m_1$ ($m_2$) is the mass of the vector meson
$V_1$ ($V_2$), 
and $b',c',x',y'$ can be obtained from 
$b,c,x,y$, respectively, with the replacement $V_1\leftrightarrow V_2$.

The calculated branching ratios for $B_s\to PP,VP,VV$ decays 
averaged over CP-conjugate modes are shown in 
Tables III-V, respectively, where the nonfactorizable effects are
treated in two different cases: (i) $\nc(LL)\neq \nc(LR)$ with the former
being fixed at the value of 2, and (ii) $\nc(LL)=\nc(LR)$. For decay
modes involving $B_s\to K^*$ or $B_s\to\phi$ transition, we apply two
different models for form factors: the BSW model [see (2.28)]
 and the light-cone sum rule approach [see (2.32)].
To compute the branching ratio, we have used the $B_s$ lifetime \cite{PDG}
\be
\tau(B_s)=(1.54\pm 0.07)\times 10^{-12}s.
\en

   From Tables III-V we see that the branching ratios for class-I and -IV modes
are stable against the variation of $\nc$ as they depend on the coefficients 
$a_1,a_4$ and $a_6$
which are $\nc$-stable. Class-V channels in general depend on the
coefficients $a_3+a_5$ and $a_7+a_9$. However, the decays
\be
\ov B_s\to\eta\pi,~\eta'\pi,~\eta\rho,~\eta'\rho,~\phi\pi,~\phi\rho
\en
do not receive any QCD penguin contributions \cite{Deshpande}.
Therefore, these six decay modes are predominantly governed by the 
electroweak penguin coefficient $a_9$, which is $\nc$-insensitive. 
A measurement of them can be utilized to fix the parameter $a_9$.
Note that their branching ratios are in general small, ranging from 
$4\times 10^{-8}$ to $0.4\times 10^{-6}$, but they could be accessible at
the future hadron colliders with large $b$ production.

   In order to see the relative importance of electroweak penguin effects
in penguin-dominated $B_s$ decays, we follow \cite{Ali3} to compute the ratio
\be
R_W=\,{{\cal B}(B_s\to h_1h_2)({\rm with}~a_7,\cdots,a_{10}=0)\over
{\cal B}(B_s\to h_1 h_2) }.
\en
Obviously, if the tree, QCD penguin and electroweak penguin amplitudes 
are of the same sign, then
$(1-R_W)$ measures the fraction of non-electroweak penguin
contributions to ${\cal B}(B_s\to h_1h_2)$. 
It is evident from Table VI that
the decays listed in (3.7) all have the same $\nc$-dependence: For
$\nc(LL)=2$, the electroweak penguin contributions account for 85\% of the
branching ratios for $B_s\to\eta\pi,\cdots,\phi\rho$, and the ratio $R_W$ is
very sensitive to $\nc$ when $\nc(LL)=\nc(LR)$. We also see that electroweak 
penguin corrections to
\be
\ov B_s\to \omega\eta,~\omega\eta',~\phi\eta,~\phi\eta',~~\omega\phi,~K
\phi,~K^*\phi,~\phi\phi,
\en
depending very sensitively on $\nc$, are in general as important as QCD 
penguin effects and even play a dominant role.
For example, about 50\% of ${\cal B}(\ov B_s\to K^0\phi)$ comes from
the electroweak penguin contributions at $\nc(LL)=2$ and $\nc(LR)=5$.

  Strictly speaking, because of variously possible interference of
the electroweak penguin amplitude with the tree and QCD penguin contributions,
$R_W$ is not the most suitable quantity for measuring the relative
importance of electroweak penguin effects. For example, it appears at
the first sight that only 21\% of ${\cal B}(B_s\to\omega\eta')$
and ${\cal B}(B_s\to\omega\phi)$  arises
from the electroweak penguins at $\nc(LL)=\nc(LR)=3$. However,   
the decay amplitudes are proportional to (see Appendix B)
\be
V_{ub}V_{us}^*a_2-V_{tb}V^*_{ts}[2(a_3+a_5)+{1\over 2}(a_7+a_9)].
\en
Since $a_2$ and $(a_3+a_5)$ are minimum at $\nc\sim 3$ (see Table I), the 
decay is obviously dominated
by the electroweak penguin transition when $\nc(LL)=\nc(LR)=3$. Numerically,
we find at the amplitude level
\be
{\rm tree:~QCD~penguin:~electroweak~penguin}=0.28:1:-2.72\,.
\en
It is clear that although $R_W=0.79$ for $\nc=3$, the decays
$B_s\to\omega\eta'$ and $B_s\to\omega\phi$ are actually
dominated by the electroweak penguin. 

The branching ratios for the class-V and -VI modes shown in (3.9) depend
strongly on the value of $\nc$. 
As pointed out in Sec.~II, the preferred values for the effective number 
of colors are $\nc(LL)\approx 2$ and $\nc(LR)\sim 5$. We believe that the
former will be confirmed soon by the forthcoming measurements of 
$B\to\pi\pi,~\pi\rho,\cdots$. However, the branching ratios for some of
the decay 
modes, e.g. $B_s\to \omega\eta,\omega\eta',\phi\eta$, become very small at the 
values of $\nc$ given by Eq.~(2.13). As suggested in \cite{Ali3}, these
decays involve large cancellation among competing amplitudes and they may
receive significant contributions from annihilation and/or final-state
interactions.

   As noted in passing, class-IV modes involve the QCD penguin 
parameters $a_4$ and
$a_6$ in the combination $a_4+Ra_6$, where $R>0$ for $B_s\to P_aP_b$, $R=0$
for $P_aV_b$ and $V_aV_b$ final states, and $R<0$ for $B_s\to V_aP_b$,
where $P_b$ or $V_b$ is factorizable under the factorization assumption.
Therefore, the decay rates of class-IV decays are expected to follow the 
pattern:
\be
\Gamma(B_s\to P_aP_b)>\Gamma(B_s\to P_aV_b)\sim\Gamma(B_s\to V_a V_b)>\Gamma
(B_s\to V_aP_b),
\en
as a consequence of various possibilities of
interference between the penguin terms characterized by the effective
coefficients $a_4$ and $a_6$. From Tables III-V, we see that
\be
&& \Gamma(\ov B_s\to K^+K^-)
>\Gamma(\ov B_s\to K^+K^{*-})\gsim \Gamma(\ov B_s\to K^{*+}K^{*-})>
\Gamma(\ov B_s\to K^{+*}K^-),   \non \\
&& \Gamma(\ov B_s\to K^0\ov K^0)
>\Gamma(\ov B_s\to K^{0}\ov K^{*0})\gsim \Gamma(\ov B_s\to K^{*0}\ov K^{*0})>
\Gamma(\ov B_s\to K^{*0}\ov K^0).
\en
Note that the pattern $\Gamma(B\to P_aV_b)>\Gamma(B\to P_aP_b)$, which is
often seen in tree-dominated decays, for example $\Gamma(\ov 
B_s\to K^+\rho^-)>\Gamma(\ov B_s\to K^+\pi^-)$,
occurs because of the larger spin phase space available to the former 
due to the existence of three different polarization states
for the vector meson. On the contrary, the hierarchy (3.13) implies
that the spin phase-space suppression of the penguin-dominated 
decay $B_s\to P_aP_b$ over
$B_s\to P_aV_b$ or $B_s\to V_aP_b$ is overcome by the constructive
interference between penguin amplitudes in the former. Recall that the
coefficient $R$ is obtained by applying equations of motion to the
hadronic matrix elements of pseudoscalar densities induced by penguin 
operators. Hence, a test of the hierarchy shown in (3.13) is important
for understanding the calculation of the penguin matrix element.
\footnote{For a direct estimate of $R$ using the perturbative QCD method
rather than the equation of motion, see \cite{Du98}.}

   Among the 39 charmless two-body decay modes of the $B_s$ meson, we find 
that only seven of them have branching ratios at the level of $10^{-5}$:
\be
\ov B_s\to K^+ K^-,~K^0\ov K^0,~\eta\eta',~\eta'\eta',~K^+\rho^-,~K^{+*}
\rho^-,~\phi\phi.
\en
It is interesting to note  that among the two-body rare decays of $B^-$ 
and $B_d$, the class-VI decays $B^-\to\eta' K^-$ and $B_d\to\eta' K^0$ have the
largest branching ratios \cite{Behrens1}:
\be
{\cal B}(B^\pm\to\eta' K^\pm) &=& \left(6.5^{+1.5}_{-1.4}\pm 0.9\right)
\times 10^{-5},  \non \\
{\cal B}(B_d\to\eta' K^0) &=& \left(4.7^{+2.7}_{-2.0}\pm 0.9\right)
\times 10^{-5}.
\en
The decay rate of $B^-\to\eta' K^-$ and $B_d\to\eta' K^0$
is large because they receive two different sets of penguin contributions
proportional to $a_4+Ra_6$ with $R>0$. By contrast, $VP,\,VV$ modes in
charm decays or bottom decays involving charmed mesons usually have 
larger branching ratios than the $PP$ mode.
Because of the strange quark content of the $B_s$, one will expect that
the decay $B_s\to\eta\eta'$ or $B_s\to\eta'\eta'$, the $B_s$ counterpart 
of $B_d\to\eta' K^0$, is the dominant two-body $B_s$ decay. Our calculation
indicates that while the branching ratio of $B_s\to\eta\eta'$ is large, 
\be
{\cal B}(B_s\to\eta\eta')\approx 2\times 10^{-5}~~{\rm for}~\nc(LL)=2,~
\nc(LR)=5,
\en
it is only slightly larger than that of other decay modes listed in (3.14), 
see Tables III-V.

   What is the role played by the intrinsic charm content of the $\eta'$ 
to the hadronic charmless $B_s$ decay ? Just as the case of $B\to\eta' K$, 
$B_s\to\etapp\eta'$ receives an internal $W$-emission contribution 
coming from the Cabibbo-allowed process $b\to c\bar c s$
followed by a conversion of the $c\bar
c$ pair into the $\eta'$ via gluon exchanges. 
Although the charm content of the $\eta'$ is {\it a priori} expected to be 
small, its contribution is potentially important because the CKM mixing
angle $V_{cb}V_{cs}^*$ is of the same order of magnitude as that of the
penguin amplitude [cf. Eqs.~(A10,A11)] and yet its effective coefficient 
$a_2$ is larger than the penguin coefficients by an order of magnitude. 
   Since $a_2$ depends strongly on $\nc(LL)$ (see Table I), the
contribution of $c\bar c\to\eta'$ is sensitive to the variation of $\nc(LL)$.
It is easy to check that the $\eta'$ charm content contributes
in the same direction as the penguin terms at $1/\nc(LL)>0.28$ where $a_2>0$,
while it contributes destructively at $1/\nc(LL)<0.28$ where $a_2$ becomes
negative. In order to explain the abnormally large branching ratio of 
$B\to\eta' K$, an enhancement from the $c\bar c\to\eta'$ mechanism is 
certainly welcome in order to improve the discrepancy between theory and 
experiment. This provides another strong support for $\nc(LL)\approx 2$.
Note that a similar mechanism explains the recent measurement of
$B^-\to\eta_c K^-$ \cite{Chan}.

   It turns out that the effect of the $c\bar c$ admixture in the $\eta'$ is 
more important for $B_s\to\eta'\eta'$ than for $B_s\to\eta\eta'$. It is clear
from Eq.~(A1) that the destructive interference between 
$X_c^{(B_s\eta,\eta')}\propto f_{\eta'}^cF_0^{B_s\eta}$ and 
$X_c^{(B_s\eta',\eta)}\propto f_{\eta}^cF_0^{B_s\eta'}$ in the decay amplitude
of $B_s\to\eta\eta'$, recalling that 
the form factors $F^{B_s\eta'}_0$ and $F_0^{B_s\eta}$ have opposite signs, 
renders the contribution of $c\bar c\to\eta'$ smaller for $B_s\to\eta\eta'$. 

   A very recent CLEO reanalysis of $B\to\eta' K$  using a 
data sample 80\% larger than in previous studies yields the preliminary
results \cite{Behrens2}:
\be
{\cal B}(B^\pm\to\eta' K^\pm) &=& \left(7.4^{+0.8}_{-1.3}\pm 1.0\right)
\times 10^{-5},  \non \\
{\cal B}(B_d\to\eta' K^0) &=& \left(5.9^{+1.8}_{-1.6}\pm 0.9\right)
\times 10^{-5},
\en
suggesting that the original measurements (3.15) were not an upward statistical
fluctuation. This result certainly favors a slightly larger
$f_{\etapp}^c$ in magnitude than that used in (2.27). In fact, 
a more sophisticated theoretical calculation gives $f_{\eta'}^c=-(12.3
\sim 18.4)$ MeV \cite{Araki}, which is consistent with all the known 
phenomenological constraints. This value of $f_{\eta'}^c$ will lead to an
enhanced decay rate for $B\to\eta' K$. Numerically, we find that for 
$\nc(LL)=2$, $\nc(LR)=5$ and $f_{\eta'}^c=-15$ MeV,
\be
{\cal B}(B_s\to\eta\eta')=2.2\times 10^{-5},  \qquad\quad {\cal B}(B_s\to
\eta'\eta')=1.8\times 10^{-5},
\en
to be compared with
\be
{\cal B}(B_s\to\eta\eta')=1.8\times 10^{-5},  \qquad\quad {\cal B}(B_s\to
\eta'\eta')=1.2\times 10^{-5},
\en
in the absence of the intrinsic charm content of the $\eta'$.

  Finally, we should point out the uncertainties associated with 
our predictions. Thus far, we
have neglected $W$-annihilation, space-like penguin diagrams, and final-state
interactions; all of them are difficult to estimate. It is argued in 
\cite{Ali3} that these effects may play an essential role for our class-V and
-VI decay modes. Other major sources of uncertainties come from the form
factors and their $q^2$ dependence,  the  running quark masses at the 
scale $m_b$, the virtual gluon's momentum in the penguin diagram, and the
values for the Wolfenstein parameters $\rho$ and $\eta$.

\section{Conclusions}
    Using the next-to-leading oder QCD-corrected effective Hamiltonian,
we have systematically studied hadronic charmless two-body decays of $B_s$
mesons within the framework of generalized factorization. Nonfactorizable
effects are parametrized in terms of $\nc(LL)$ and $\nc(LR)$, the
effective numbers of colors arising from $(V-A)(V-A)$ and $(V-A)(V+A)$
4-quark operators, respectively. The branching ratios are calculated as 
a function of $\nc(LR)$ with two different considerations for $\nc(LL)$:
(i) $\nc(LL)$ being fixed at the value of 2, 
and (ii) $\nc(LL)=\nc(LR)$. Depending on the sensitivity of the effective
coefficients $a_i^{\rm eff}$ on $\nc$, we have classified the tree and penguin 
transitions into six different classes. Our results are:
\begin{enumerate}

\item The decays $\ov B_s\to\eta\pi,~\eta'\pi,~\eta\rho,~\eta'\rho,~\phi
\pi,~\phi\rho$ receive contributions only from the tree and electroweak 
penguin diagrams and are completely dominated by the latter. A measurement 
of them can 
be utilized to fix the effective electroweak penguin parameter $a_9$.
For $\nc(LL)=2$, we found that electroweak penguin contributions account
for 85\% of their decay rates. Their branching ratios, though small [\,in
the range of $(0.4-4.0)\times 10^{-7}$], could be accessible at hadron 
colliders with large $b$ production. 

\item For class-V and -VI penguin-dominated modes:
$\ov B_s\to \omega\eta,~\omega\eta',~\phi\eta,~\omega
\eta,~K\phi,~K^*\phi,~\phi\phi$, electroweak penguin corrections, depending
strongly on $\nc$, are as significant as QCD penguin effects and can even
play a dominant role.

\item Current experimental information on $B^-\to\omega\pi^-$ and 
$B^0\to\pi^+\pi^-$ favors a small $\nc(LL)$, that is, $\nc(LL)\approx 2$, which
is also consistent with the nonfactorizable term extracted from $B\to (D,D^*) 
(\pi,\rho)$ decays, $\nc(B\to D\pi)\approx 2$. 
We have argued that the 
preferred value for the effective number of colors $\nc(LR)$ is 
$\nc(LR)\sim 5$.

\item Because of various possibilities of interference between the penguin 
amplitudes governed by the QCD penguin parameters $a_4$ and $a_6$,
the decay rates of class-IV decays follow the pattern:
$\Gamma(\ov B_s\to P_aP_b)>\Gamma(\ov B_s\to P_aV_b)\sim\Gamma(\ov 
B_s\to V_a V_b)>\Gamma(\ov B_s\to V_aP_b)$, where $P_a=K^+,~P_b=K^-$ or 
$P_a=K^0,~P_b=\ov K^0$. A test 
of this hierarchy is important to probe the penguin mechanism.

\item The decay $B\to\eta' K$ is known to have the largest branching ratios
in the two-body hadronic charmless $B^-$ and $B_d$ decays. Its analogue in 
the $B_s$ system, namely $B_s\to\eta\eta'$ has a branching ratio of order 
$2\times 10^{-5}$, but it is only slightly larger than that of 
$\eta'\eta',K^{*+}\rho^-,K^+K^-,K^0\ov K^0$ decay modes, which have the
branching ratios of order $10^{-5}$.

\item The recent CLEO reanalysis of $B\to\eta' K$ favors a slightly large
decay constant $f_{\eta'}^c$. Using $f_{\eta'}^c=-15$ MeV, which is
consistent with all the known theoretical and phenomenological constraints,
we found that the intrinsic charm content of the $\eta'$ is important for 
$B_s\to\eta'\eta'$, but less significant for $B_s\to\eta\eta'$. 

\end{enumerate}

\acknowledgements

We are grateful to Cai-Dian L\"u and Kwei-Chou Yang for reading the
manuscript and for their useful comments.
This work is supported in part by the National Science Council of
the Republic of China under Grants Nos. NSC88-2112-M001-006 and 
NSC88-2112-M006-013.

\vskip 1cm
\centerline{\bf APPENDIX}
\vskip 0.3cm
\noindent { \bf A. The $\ov B_s \to P P$ decay amplitudes }
\renewcommand{\thesection}{\Alph{section}}
\renewcommand{\theequation}{\thesection\arabic{equation}}
\setcounter{equation}{0}
\setcounter{section}{1}
\bigskip

For $\ov B_s\to PP$ decays, we use
$X^{(B_sP_1,P_2)}$ to denote the factorizable amplitude with the
meson $P_2$ being factored out. Explicitly,
\begin{equation}
X^{(B_s P_1,P_2)} \equiv \la P_2|
(\bar{q}_2 q_3)_\vma|0\ra\la P_1|(\bar
{q}_1b)_\vma|\ov B_s
\ra=if_{P_2}(m_{B_s}^2-m^2_{P_1}) F_0^{B_s P_1}(m_{P_2}^2).
\end{equation}
For a neutral $P_1$ with the quark content $N(\bar qq+\cdots)$, where
$N$ is a normalization constant, 
\begin{equation}
X^{(B_s P_1,P_2)}_q \equiv \la P_2|
(\bar{q} q)_\vma|0\ra\la P_1|(\bar{q}_1b)_\vma|\ov B_s
\ra=if_{P_2}^q(m_{B_s}^2-m^2_{P_1}) F_0^{B_s P_1}(m_{P_2}^2).
\end{equation}
As an example, the factorizable amplitudes $X^{(B_s\eta',K)}$ and $X_q^{(B_sK,
\eta')}$ of the decay $\ov B_s\to\ov K^0\eta'$ read
\be
X^{(B_s\eta',K)}
&=& \la \overline{K}^0|(\bar sd)_\vma|0\ra\la
\eta'|(\bar db)_\vma|\ov B_s\ra
= if_K(m_{B_s}^2-m^2_{\eta'})F_0^{B_s\eta'}(m_K^2),   \non \\
X^{(B_s K,\eta')}_q
&=& \la \eta'|(\bar qq)_\vma|0\ra \la\overline{K}^0|(\bar sb)_
\vma|\ov B_s\ra
= if_{\eta'}^q(m_{B_s}^2-m^2_K)F_0^{B_s K}(m_{\eta'}^2).
\en
For simplicity, $W$-annihilation, space-like penguins and final-state
interactions are not included in the decay amplitudes given below.

\bigskip
\noindent 1.~~$b\to d$ processes:

\medskip
\be
A(\ov B_s \to K^+ \pi^-) &=& {G_F\over\sqrt{2}}\Bigg\{
V_{ub}V_{ud}^* ~a_1 
- V_{tb}V_{td}^* \Bigg[ a_4+ a_{10}+ \non \\
&+& 2(a_6+a_8){m_{\pi}^2 \over
(m_u+m_d)(m_b-m_u)}\Bigg] \Bigg\}
X^{(B_s K^+,\pi^- )},
\en

\medskip
\be
A(\ov B_s \to K^0 \pi^0) &=& {G_F\over\sqrt{2}}\Bigg\{
V_{ub}V_{ud}^* ~a_2 
- V_{tb}V_{td}^* \Bigg[ - a_4+{3\over 2}(-a_7+a_9)+{1\over
2}a_{10}\non \\
&-& 2(a_6-{1\over 2}a_8){m_{\pi}^2 \over
(m_d+m_d)(m_b-m_d)}\Bigg] \Bigg\}
X^{(B_s K^0,\pi^0 )}_u,
\en

\medskip
\be
A(\ov B_s\to K^0 \eta^{(')})
&=& {G_F\over\sqrt{2}}\Bigg\{
V_{ub}V_{ud}^*\,a_2X^{({B_s}K,\eta^{(')})}_u
+V_{cb}V_{cd}^*\,a_2 X^{({B_s}K,\eta^{(')})}_c \non \\
&-& V_{tb}V_{td}^*\Bigg[ \left(a_4-{1\over 2}a_{10} 
+2(a_6-{1\over 2}a_8){m_K^2\over
(m_s+m_d)(m_b-m_d)}\right)X^{({B_s}\eta^{(')},K)}   \non \\
&+& \left(a_3-a_5-a_7+a_9\right)X^{({B_s}K,\eta^{(')})}_u   
+ (a_3-a_5+{1 \over 2}a_7-{1\over 2}a_9)X^{({B_s}K,\eta^{(')})}_s \non \\
&+& (a_3-a_5-a_7+a_9)X_c^{({B_s}K,\eta^{(')})}   \non \\
&+& \Bigg(a_3+a_4-a_5+{1\over 2}a_7-{1\over 2}a_9-{1\over 2}a_{10} \non\\
&+& (a_6-{1\over 2}a_8){m^2_{\eta^{(')}}\over
m_s(m_b-m_s)}\left({{f_{\eta^{(')}}^s
\over f_{\etapp}^u} -1}\right) 
r_{\eta^{'}}\Bigg)X^{({B_s}K,\eta^{(')})}_d \Bigg]\Bigg\}.
\en

\medskip
\noindent 2.~~$b\to s$ processes:

\medskip
\be
A(\ov B_s \to K^+ K^-) &=& {G_F\over\sqrt{2}}\Bigg\{
V_{ub}V_{us}^* ~a_1 
- V_{tb}V_{ts}^* \Bigg[ a_4+ a_{10}+ \non \\
&+& 2(a_6+a_8){m_{K}^2 \over
(m_u+m_s)(m_b-m_u)}\Bigg] \Bigg\}
X^{(B_s K^+,K^- )},
\en

\medskip
\be
A(\ov B_s\to\pi^0\eta^{(')} ) = {G_F\over\sqrt{2}}\Bigg\{
V_{ub}V_{us}^* \,a_2 
- V_{tb}V_{ts}^*\Bigg[ {3\over 2}(-a_7+a_9)\Bigg]\Bigg\}
X^{(B_s\eta^{(')},\pi^0)}_u,
\en
where
\be
X_u^{(B_s\eta^{(')},\pi^0)} &\equiv& \la \pi^0|(\bar
uu)_\vma|0\ra\la\eta^{(')}|(\bar
sb)_\vma|\ov B_s\ra
= i{{f_\pi} \over
\sqrt{2}}(m_{B_s}^2-m^2_{\eta^{(')}})F_0^{B_s\eta^{(')}}(m_\pi^2),
\en

\medskip
\be
A(\ov B_s\to\eta \eta')
&=& {G_F\over\sqrt{2}}\Bigg\{
V_{ub}V_{us}^*\,a_2\left(X^{(B_s\eta,\eta')}_u
+X^{(B_s\eta',\eta)}_u\right)
+V_{cb}V_{cs}^*\,a_2\left(X^{(B_s\eta,\eta')}_c
+X^{(B_s\eta',\eta)}_c\right)
\non \\
&-& V_{tb}V_{ts}^*\Bigg[
 \Bigg(a_3+a_4-a_5+{1\over 2}a_7-{1\over 2}a_9-{1\over 2}a_{10} \non \\
&+&(a_6-{1\over 2}a_8){m^2_{\eta'}\over 
m_s(m_b-m_s)}\left(1-{f_{\eta'}^u
\over f_{\eta'}^s}\right)\Bigg)X^{(B_s\eta,\eta')}_s
\non \\
&+& \left(2a_3-2a_5-{1\over 2}a_7+{1 
\over 2}a_9\right)X^{(B_s\eta,\eta')}_u
+(a_3-a_5-a_7+a_9)X_c^{(B_s\eta,\eta')} \non \\
&+& \Bigg(a_3+a_4-a_5+{1\over 2}a_7-{1\over 2}a_9-{1\over 2}a_{10} \non \\
&+& (a_6-{1\over 2}a_8){m^2_{\eta}\over 
m_s(m_b-m_s)}\left(1-{f_{\eta}^u  
\over f_{\eta}^s}\right)\Bigg)X^{(B_s\eta',\eta)}_s
\non \\
&+& \left(2a_3-2a_5-{1\over 
2}a_7+{1\over 2}a_9\right)X^{(B_s\eta',\eta)}_u
+(a_3-a_5-a_7+a_9)X_c^{(B_s\eta',\eta)}
\Bigg]\Bigg\},
\en

\medskip
\be
A(\ov B_s\to\eta' \eta')
&=& {G_F\over\sqrt{2}}\,2\Bigg\{
V_{ub}V_{us}^* \,a_2X^{(B_s\eta',\eta')}_u
+V_{cb}V_{cs}^* \,a_2X^{(B_s\eta',\eta')}_c \non \\
&-& V_{tb}V_{ts}^*\Bigg[
 \Bigg(a_3+a_4-a_5+{1\over 2}a_7-{1\over 2}a_9-{1\over 2}a_{10} \non \\
&+&(a_6-{1\over 2}a_8){m^2_{\eta'}\over m_s(m_b+m_s)}\left(1-{f_{\eta'}^u
\over f_{\eta'}^s}\right)\Bigg)X^{(B_s\eta',\eta')}_s   \non \\
&+& \left(2a_3-2a_5-\half a_7+\half a_9\right)X^{(B_s\eta',\eta')}_u
+(a_3-a_5-a_7+a_9)X_c^{(B_s\eta',\eta')}
\Bigg]\Bigg\}.
\en

The amplitude of $\ov B_s\to\eta\eta$ is obtained from $A(\ov B_s\to 
\eta'\eta')$ by the replacement $\eta'\to\eta$.

\medskip
\noindent 3.~~pure penguin process:
\medskip
\be
A(\ov B_s \to K^0 \ov{K}^0) &=& {G_F\over\sqrt{2}}\Bigg\{
- V_{tb}V_{ts}^* \Bigg[ a_4-{1\over 2} a_{10} \non \\
&+& 2(a_6-{1 \over 2}a_8){m_{K}^2 \over
(m_s+m_d)(m_b-m_d)}\Bigg] \Bigg\}
X^{(B_s K^0,\ov{K}^0 )}.
\en
\bigskip
\medskip

\bigskip
\noindent{ \bf B. The $\ov B_s \to V P$ decay amplitudes}
\bigskip
\setcounter{equation}{0}
\setcounter{section}{2}

The factorizable amplitudes of $\ov B_s\to VP$ decays have the form:
\be
X^{(B_sP,V)} &\equiv & \la V|
(\bar{q}_2 q_3)_\vma|0\ra\la P|(\bar{q}_1b)_\vma|\ov B_s
\ra=2f_V\,m_V F_1^{B_s P}(m_{V}^2)(\vp\cdot p_{_{B_s}}),   \non \\
X^{(B_sV,P)} &\equiv & \la P |
(\bar{q}_2 q_3)_\vma|0\ra\la V|(\bar{q}_1b)_\vma|\ov B_s
\ra=2f_P\,m_V A_0^{B_s V}(m_{P}^2)(\vp\cdot p_{_{B_s}}).
\en
For example, the factorizable terms $X^{(B_s\eta',K^*)}$ and 
$X^{(B_sK^*,\eta')}_q$ of $\ov B_s\to K^*\eta'$ decay are given by
\be 
X^{(B_s \eta',K^*)} &\equiv& \la K^{*0}|(\bar su)_\vma|0\ra\la
\eta'|(\bar ub)_\vma|\ov B_s\ra
= 2f_{K^*}m_{K^*}F_1^{B_s\eta'} (m_{K^*}^2)(\vp\cdot p_{_{B_s}}),   \non \\
X^{(B_s K^*,\eta')}_q
&\equiv& \la \eta'|(\bar qq)_\vma|0\ra\la 
K^{*0}|(\bar sb)_\vma|\ov B_s \ra
= 2f_{\eta'}^q m_{K^*}A_0^{B_sK^*}(m_{\eta'}^2)(\vp\cdot p_{_{B_s}}).
\en

\medskip
\noindent 1.~~$b\to d$ processes:

\medskip
\be
A(\ov B_s \to K^{+\ast} \pi^{-}) &=& {G_F\over\sqrt{2}}\Bigg\{
V_{ub}V_{ud}^* ~a_1 
- V_{tb}V_{td}^*\Bigg[  a_4+
a_{10}\non \\
&-& 2(a_6 + a_8){m_{\pi }^2\over
(m_u + m_d)(m_b + m_u)}\Bigg]\Bigg\}
X^{(B_s K^{+\ast},\pi^{-})}, 
\en

\medskip
\be
A(\ov B_s \to K^+ \rho^{-}) &=& {G_F\over\sqrt{2}}\Bigg\{
V_{ub}V_{ud}^* ~a_1 
- V_{tb}V_{td}^*\Big( a_4+a_{10} \Big) \Bigg\}
X^{(B_s K^+,\rho^{-})},
\en

\medskip
\be
A(\ov B_s \to K^{0\ast} \pi^{0}) &=& {G_F\over\sqrt{2}}\Bigg\{
V_{ub}V_{ud}^* ~a_2 
- V_{tb}V_{td}^*\Bigg[ -a_4 - {3\over 2} a_7
+ {3\over 2}a_9 + {1\over 2} a_{10}\non \\
&+& 2(a_6 -{1\over 2} a_8){m_{\pi }^2\over
2m_d(m_b + m_d)}\Bigg]\Bigg\}
X^{(B_s K^{0\ast},\pi^{0})}_u, 
\en

\medskip
\be
A(\ov B_s \to K^{0} \rho^{0}) &=& {G_F\over\sqrt{2}}\Bigg\{
V_{ub}V_{ud}^* ~a_2 
- V_{tb}V_{td}^*\Big( -a_4 + {3\over 2} a_7
+ {3\over 2}a_9 + {1\over 2} a_{10}\Big)\Bigg\}
X^{(B_s K^{0},\rho^{0})}_u,\non \\ 
\en

\medskip
\be
A(\ov B_s \to K^{0} \omega) &=& {G_F\over\sqrt{2}}\Bigg\{
V_{ub}V_{ud}^* ~a_2 
- V_{tb}V_{td}^*\Big(2a_3 +a_4+2a_5  \non \\
&+& {1\over 2} a_7
+ {1\over 2}a_9 - {1\over 2} a_{10}
\Big)\Bigg\}X^{(B_s K^{0},\omega)}_u, 
\en

\medskip
\be
A(\ov B_s \to \ov K^{*0}\etapp) &=& {G_F\over\sqrt{2}}\Bigg\{
V_{ub}V_{ud}^*\,a_2X^{(B_s K^*,\eta^{'})}_u 
+V_{cb}V_{cd}^*\,a_2X^{(B_sK^*,\eta^{'})}_c \non \\
&-& V_{tb}V_{td}^*\Bigg[
(a_4-{1\over 2}a_{10})X^{(B_s\eta^{'},K^*)}
+ \Bigg(a_3+a_4-a_5+{1\over 2}a_7-{1\over 2}a_9-{1\over 2}a_{10} \non \\
&-& (a_6-{1\over 2}a_8){m^2_{\eta^{'}}\over
m_s(m_b+m_s)}
\left({{f_{\eta^{'}}^s \over f_{\eta^{'}}^u} -1}\right) r_{\eta^{'}} 
\Bigg)X^{(B_sK^*,\eta^{'})}_d \non \\  
&+& \left(a_3-a_5-a_7+a_9\right)X^{(B_sK^*,\eta^{(')})}_u
+ (a_3-a_5+{1\over 2}a_7-{1\over 2}a_9)X^{(B_sK^*,\eta^{(')})}_s
\non \\
&+&(a_3-a_5-a_7+a_9)X_c^{(B_s K^*,\eta^{(')})}
\Bigg]\Bigg\}.
\en

\medskip
\noindent 2.~~$b\to s$ processes:

\medskip
\be
A(\ov B_s \to K^+ K^{-\ast}) &=& {G_F\over\sqrt{2}}\Bigg\{
V_{ub}V_{us}^* ~a_1 
- V_{tb}V_{ts}^*\Big( a_4+ a_{10} \Big) \Bigg\}
X^{(B_s K^+,K^{-\ast})},
\en

\medskip
\be
A(\ov B_s \to K^{+\ast} K^{-}) &=& {G_F\over\sqrt{2}}\Bigg\{
V_{ub}V_{us}^* ~a_1 
- V_{tb}V_{ts}^*\Bigg[  a_4+
a_{10}\non \\
&-& 2(a_6 + a_8){m_{K }^2\over
(m_s + m_u)(m_b + m_u)}\Bigg]\Bigg\}
X^{(B_s K^{+\ast},K^{-})}, 
\en

\medskip
\be
A(\ov B_s \to \rho^{0} \eta^{(')}) = {G_F\over\sqrt{2}}\Bigg\{
V_{ub}V_{us}^* ~a_2 - V_{tb} V^{*}_{ts}\Bigg[ {3\over 2} (a_7 + a_9)\Bigg]
\Bigg\}X^{(B_s \eta^{(')},\rho^{0})}_u,
\en

\medskip
\be
A(\ov B_s \to \omega \eta^{(')} ) = {G_F\over\sqrt{2}}\Bigg\{
V_{ub}V_{us}^* \,a_2
- V_{tb}V_{ts}^*\Bigg[ 2(a_3+a_5)+{1\over 2}(a_7+a_9)\Bigg]\Bigg\}
X^{(B_s\eta^{(')},\omega)}_u,
\en
where
\be
X_u^{(B_s\eta^{(')},\omega)} &\equiv& \la \omega|(\bar
uu)_\vma|0\ra\la\eta^{(')}|(\bar
sb)_\vma|\ov B_s\ra
= {\sqrt{2}} {f_\omega} m_{\omega}F_1^{B\eta^{(')}}(m_\omega^2)(\vp\cdot
p_{_{B_s}}),
\en

\medskip
\be
A(\ov B_s\to \pi^0 \phi ) = {G_F\over\sqrt{2}}\Bigg\{
V_{ub}V_{us}^*\, a_2
- V_{tb}V_{ts}^*\Bigg[ {3  \over 2}(-a_7+a_9)\Bigg]\Bigg\}
X^{(B_s \phi,\pi^0)}_u,
\en

\medskip
\be
A(\ov B_s\to\phi \eta^{(')})
&=& {G_F\over\sqrt{2}}\Bigg\{
V_{ub}V_{us}^* \,a_2X^{(B_s\phi,\eta^{(')})}_u
+V_{cb}V_{cs}^*\,a_2X^{(B_s\phi,\eta^{(')})}_c \non \\
&-& V_{tb}V_{ts}^*\Bigg[
 \Bigg(a_3+a_4-a_5+{1\over 2}a_7-{1\over 2}a_9-{1\over 2}a_{10} \non \\
&-&(a_6-{1\over 2}a_8){m^2_{\eta^{(')}}\over
m_s(m_b+m_s)}\left(1-{f_{\eta^{(')}}^u
\over f_{\eta^{(')}}^s}\right)\Bigg)X^{(B_s\phi,\eta^{(')})}_s   \non \\
&+& \left(2a_3-2a_5-{1\over 2}a_7+{1\over 
2}a_9\right)X^{(B_s\phi,\eta^{(')})}_u
+(a_3-a_5-a_7+a_9)X_c^{(B_s\phi,\eta^{(')})} \non \\
&+& \Bigg(a_3+a_4+a_5-{1\over 2}a_7-{1\over 2}a_9-{1\over 2}a_{10} 
\Bigg)X^{(B_s\eta^{(')},\phi)}
\Bigg]\Bigg\}.
\en

\medskip
\noindent 3.~~pure penguin processes:
\medskip
\be
A(\ov B_s \to K^0 \ov{K}^{0\ast}) &=&
- {G_F\over\sqrt{2}}\,V_{tb}V_{ts}^*\Big( a_4-{1\over 2}
a_{10} \Big) X^{(B_s K^0,\bar{K}^{0\ast})},
\en

\medskip
\be
A(\ov B_s \to K^{0\ast} \ov{K}^{0}) &=& 
- {G_F\over\sqrt{2}}\,V_{tb}V_{ts}^*\Bigg[  a_4 - {1\over 2}
a_{10}   \non \\
&-& 2(a_6 - {1\over 2} a_8){m_{K}^2\over
(m_s + m_d)(m_b + m_d)}\Bigg] X^{(B_s K^{0\ast},\bar{K}^{0})}, 
\en

\medskip
\be
A(\ov B_s \to K^{0} \phi) &=&
- {G_F\over\sqrt{2}}\,V_{tb}V_{td}^*\Bigg\{ \Bigg[a_3 + a_5 -{1\over 2}
(a_7 + a_9)\Bigg] X^{(B_s K^{0},\phi)}\non \\
&+& \Bigg[a_4 - {1\over 2} a_{10}
- 2(a_6 -{1\over 2} a_8){m_{K }^2\over
(m_s + m_d)(m_b + m_d)}\Bigg]
X^{(B_s \phi,K^{0})}\Bigg\}.\non \\ 
\en

\medskip
\medskip
\noindent{ \bf C. The $\ov B_s \to V V$ decay amplitudes }
\setcounter{equation}{0}
\setcounter{section}{3}
\bigskip

The factorizable amplitude of $B_s\to VV$ decays has the form:
\be
X^{(B_sV_1,V_2)} &=& if_{V_2}m_{_{V_2}}\Bigg[ (\vp^*_1\cdot\vp^*_2)
(m_{B_s}+m_{V_1})A_1^{B_sV_1}(m_{V_2}^2)  \non \\
&-& (\vp^*_1\cdot p_{_{B_s}})(\vp^*_2
\cdot p_{_{B_s}}){2A_2^{B_sV_1}(m_{V_2}^2)\over (m_{B_s}+m_{V_1}) }  \non \\
&+& i\epsilon_{\mu\nu\alpha\beta}\vp^{*\mu}_2\vp^{*\nu}_1p^\alpha_{_{B_s}}
p^\beta_1\,{2V^{B_sV_1}(m_{V_2}^2)\over (m_{B_s}+m_{V_1}) }\Bigg].
\en

\bigskip
\noindent 1.~~$b\to d$ processes:
\medskip
\be
A(\ov B_s \to K^{+\ast} \rho^{-}) &=& {G_F\over\sqrt{2}}\Bigg\{
V_{ub}V_{ud}^* ~a_1 
- V_{tb}V_{td}^*\Big( a_4+
a_{10} \Big) \Bigg\}
X^{(B_s K^{+\ast},\rho^{-})},
\en

\medskip
\be
A(\ov B_s \to K^{0\ast} \rho^{0}) &=& {G_F\over\sqrt{2}}\Bigg\{
V_{ub}V_{ud}^* ~a_2 
- V_{tb}V_{td}^*\Big( -a_4 + {3\over 2} a_7
+ {3\over 2}a_9 + {1\over 2} a_{10}
\Big)\Bigg\}X^{(B_s K^{0\ast},\rho^{0})}_u, 
\en

\medskip
\be
A(\ov B_s \to K^{0\ast} \omega) &=& {G_F\over\sqrt{2}}\Bigg\{
V_{ub}V_{ud}^* ~a_2 
- V_{tb}V_{td}^*\Big( 2a_3 +a_4+2a_5 \non \\
&+& {1\over 2} a_7
+ {1\over 2}a_9 - {1\over 2} a_{10}
\Big) \Bigg\}X^{(B_s K^{0\ast},\omega)}_u.
\en
\medskip

\noindent 2.~~$b\to s$ processes:

\medskip
\be
A(\ov B_s \to K^{+\ast} K^{-\ast}) &=& {G_F\over\sqrt{2}}\Bigg\{
V_{ub}V_{us}^* ~a_1 
- V_{tb}V_{ts}^*\Big( a_4+
a_{10} \Big) \Bigg\}
X^{(B_s K^{+\ast},K^{-\ast})},
\en

\medskip
\be
A(\ov B_s \to \rho^{0} \phi) &=& {G_F\over\sqrt{2}}\Bigg\{
V_{ub}V_{us}^* ~a_2 
- V_{tb}V_{ts}^* \Bigg[{3\over 2} (a_7
+ a_9)\Bigg]\Bigg\}X^{(B_s \phi,\rho^{0})}_u, 
\en

\medskip
\be
A(\ov B_s \to \omega \phi) &=& {G_F\over\sqrt{2}}\Bigg\{
V_{ub}V_{us}^* ~a_2 
- V_{tb}V_{ts}^* \Bigg[2(a_3 + a_5) + {1\over 2} (a_7
+ a_9)
\Bigg]\Bigg\}
X^{(B_s \phi,\omega)}_u.
\en

\medskip
\noindent 3.~~pure penguin processes:

\medskip
\be
A(\ov B_s \to K^{0\ast} \ov{K}^{0\ast}) &=&
- {G_F\over\sqrt{2}}\,V_{tb}V_{ts}^*\Big( a_4-{1\over 2}
a_{10} \Big) 
X^{(B_s K^{0\ast},\ov{K}^{0\ast})},
\en

\medskip
\be
A(\ov B_s \to K^{0\ast} \phi) &=&
- {G_F\over\sqrt{2}}\,V_{tb}V_{td}^*\Bigg\{ \Bigg[a_3 + a_5 -{1\over 2}
(a_7 + a_9)\Bigg] X^{(B_s K^{0\ast},\phi)}   \non\\
&+&  \Big(a_4 - {1\over 2} a_{10}\Big)X^{(B_s \phi,K^{0\ast})}\Bigg\}, 
\en

\medskip
\be
A(\ov B_s \to \phi \phi) &=&
- {G_F\over\sqrt{2}}\,V_{tb}V_{ts}^*\, 2\Bigg[ a_3 + a_4 +a_5 -{1\over 2}
(a_7 + a_9 + a_{10})\Bigg]
X^{(B_s \phi, \phi)}.
\en


\vskip 0.4cm
\begin{table}[ht]
{\small Table III. Branching ratios (in units of $10^{-6}$) averaged over 
CP-conjugate modes for charmless $\ov B_s\to PP$ decays. 
Predictions are for $k^2=m_b^2/2$, $\eta=0.34,~\rho=0.16$, and 
$\nc(LR)=2,3,5,\infty$ with $\nc(LL)$ being fixed to be 2 in the first 
case and treated to be the same as $\nc(LR)$ in the second case. We
use the BSW model for form factors [see (2.28)].}
\begin{center}
\begin{tabular}{l c c c c c c c c c } 
 &  & \multicolumn{4}{c}{$\nc(LL)=2$}  
 &   \multicolumn{4}{c}{$\nc(LL)=\nc(LR)$}  \\ \cline{3-6} \cline{7-10}
\raisebox{2.0ex}[0cm][0cm]{Decay} & \raisebox{2.0ex}[0cm][0cm]{Class} & 
2 & 3 & 5 & $\infty$ & 2 & 3 & 5
& $\infty$  \\ 
\hline 
$\ov B_s\to K^+\pi^-$ & I & 6.64 & 6.66 & 6.67 & 6.70 & 6.64 & 7.38 & 8.01 & 
8.99 \\
$\ov B_s\to K^0\pi^0$ & II & 0.24 & 0.24 & 0.25 & 0.25 & 0.24 & 0.08 & 0.12 
& 0.46 \\
$\ov  B_s\to K^+ K^-$ & IV & 9.88 & 10.9 & 10.9 & 11.6 & 9.88 & 10.9 & 11.7 & 
12.9 \\
$\ov B_s\to K^0\ov K^0$ & IV & 10.3 & 10.9 & 11.4 & 12.1 & 10.3 & 12.0 & 
13.5 & 15.8 \\ 
$\ov B_s\to\pi^0\eta' $ & V & 0.04 & 0.04 & 0.04 & 0.04 & 0.04 & 0.05 & 
0.06 & 0.09 \\
$\ov B_s\to\pi^0\eta $ & V & 0.04 & 0.04 & 0.04 & 0.04 & 0.04 & 0.05 & 
0.06 & 0.09 \\
$\ov B_s\to K^0\eta'$ & VI & 0.63 & 0.86 & 1.06 & 1.42 & 0.63 & 0.54 & 0.57 
& 0.76 \\
$\ov B_s\to K^0\eta$ & VI & 0.81 & 0.84 & 0.87 & 0.91 & 0.81 & 0.82 & 0.96 & 
1.39 \\
$\ov B_s\to\eta \eta'$ & VI & 12.5 & 16.3 & 19.6 & 25.3 & 12.5 & 14.4 & 
15.9 & 18.5 \\
$\ov B_s\to\eta'\eta'$ & VI & 6.28 & 10.3 & 14.3 & 21.4 & 6.28 & 6.80 & 
7.23 & 7.91 \\
$\ov B_s\to\eta \eta$ & VI & 5.30 & 4.80 & 4.41 & 3.89 & 5.30 & 6.23 & 
7.05 & 8.37 \\
\end{tabular}
\end{center} 
\end{table} 
\vskip 0.4cm

\vskip 0.4cm
\begin{table}[ht]
{\small Table IV. Branching ratios (in units of $10^{-6}$) averaged over
CP-conjugate modes for charmless $\ov B_s\to VP$ decays. 
Predictions are for $k^2=m_b^2/2$, $\eta=0.34,~\rho=0.16$, and 
$\nc(LR)=2,3,5,\infty$ with $\nc(LL)$ being fixed to be 2 in the first 
case and treated to be the same as $\nc(LR)$ in the second case. For decay
modes involving the $B_s\to K^*$ or $B_s\to\phi$ transition, we use two
different models for form factors: the BSW model \cite{BSW87} (the upper
entry) and the light-cone sum rule approach \cite{Ball} (the lower entry).}
\begin{center}
\begin{tabular}{l c c c c c c c c c } 
 & &  \multicolumn{4}{c}{$\nc(LL)=2$}  
 &  \multicolumn{4}{c}{$\nc(LL)=\nc(LR)$}  \\ \cline{3-6} \cline{7-10}
\raisebox{2.0ex}[0cm][0cm]{Decay} & \raisebox{2.0ex}[0cm][0cm]{Class} 
& 2 & 3 & 5 & $\infty$ & 2 & 3 & 5
& $\infty$  \\ 
\hline
$\ov B_s\to K^{*+}\pi^-$ & I & 4.30 & 4.30 & 4.30 & 4.30 & 4.30 & 4.79 & 5.20 
& 5.84 \\
& & 4.98 & 4.98 & 4.98 & 4.98 & 4.98 & 5.55 & 6.02 & 6.76 \\
$\ov B_s\to K^+\rho^-$ & I & 17.2 & 17.2 & 17.2 & 17.2 & 17.2 & 19.2 & 20.8 
& 23.4 \\
$\ov B_s\to K^{0*}\pi^0$ & II & 0.14 & 0.14 & 0.14 & 0.15 & 0.14 & 0.01 & 
0.02 & 0.23 \\
& & 0.17 & 0.17 & 0.17 & 0.17 & 0.17 & 0.01 & 0.02 & 0.27 \\
$\ov B_s\to K^0\rho^0$ & II & 0.55 & 0.55 & 0.55 & 0.55 & 0.55 & 0.06 & 0.13 & 
1.00 \\
$\ov B_s\to K^{*0}\eta'$ & II,VI & 0.10 & 0.14 & 0.18 & 0.26 & 0.10 & 0.04 
& 0.04 & 0.15  \\
& & 0.11 & 0.15 & 0.20 & 0.29 & 0.11 & 0.03 & 0.04 & 0.17 \\ 
$\ov B_s\to K^{*0}\eta$ & II,VI & 0.18 & 0.18 & 0.18 & 0.17 & 0.18 & 0.13 & 
0.17 & 0.39 \\ 
& & 0.20 & 0.19 & 0.19 & 0.19 & 0.20 & 0.13 & 0.18 & 0.42 \\
$\ov B_s\to K^0\omega$ & II,VI & 0.71 & 0.60 & 0.53 & 0.46 & 0.71 & 0.11 & 
0.07 & 0.77 \\ 
$\ov B_s\to K^{+*}K^-$ & IV & 0.68 & 0.78 & 0.87 & 1.01 & 0.68 & 0.75 & 0.80 & 
0.88 \\
& & 0.79 & 0.90 & 1.00 & 1.16 & 0.79 & 0.86 & 0.92 & 1.02 \\
$\ov B_s\to K^{0*}\overline K^0$ & IV & 0.26 & 0.34 & 0.41 & 0.53 & 0.26 & 
0.20 & 0.15 & 0.10 \\
& & 0.31 & 0.40 & 0.48 & 0.62 & 0.31 & 0.23 & 0.18 & 0.11 \\
$\ov B_s\to K^+K^{-*}$ & IV & 3.40 & 3.40 & 3.40 & 3.56 & 3.40 & 3.77 & 4.07 & 
4.55 \\
$\ov B_s\to K^0\overline K^{0*}$ & IV & 3.28 & 3.28 & 3.28 & 3.28 & 3.28 & 
4.15 & 4.92 & 6.21  \\
$\ov  B_s\to \pi^0\phi$ & V & 0.18 & 0.18 & 0.18 & 0.17 & 0.18 & 0.22 & 
0.27 & 0.40 \\
& & 0.35 & 0.34 & 0.34 & 0.33 & 0.35 & 0.42 & 0.53 & 0.78 \\
$\ov B_s\to\rho\eta' $ & V & 0.11 & 0.11 & 0.11 & 0.11 & 0.11 & 0.13 & 0.17 
& 0.26 \\
$\ov B_s\to\rho\eta $ & V & 0.11 & 0.11 & 0.11 & 0.11 & 0.11 & 0.14 & 0.18 & 
0.26 \\
$\ov B_s\to \omega \eta' $ & V & 0.79 & 0.18 & 0.01 & 0.31 & 0.79 & 0.004 & 
0.36 & 2.52 \\
$\ov B_s\to \omega \eta $ & V & 0.80 & 0.18 & 0.01 & 0.31 & 0.80 & 0.004 & 
0.36 & 2.56 \\
$\ov B_s\to\phi \eta'$ & VI & 1.06 & 1.18 & 1.28 & 1.45 & 1.06 & 0.27 & 
0.22 & 1.11 \\
& & 0.55 & 0.86 & 1.20 & 1.86 & 0.55 & 0.31 & 0.75 & 2.45 \\
$\ov B_s\to\phi \eta$ & VI & 2.03 & 0.79 & 0.25 & 0.20 & 2.03 & 0.91 & 
0.34 & 0.04  \\
& & 1.43 & 0.41 & 0.15 & 0.69 & 1.43 & 0.58 & 0.19 & 0.09 \\
$\ov B_s\to K^0\phi$ & VI & 0.002 & 0.01 & 0.04 & 0.13 & 0.002 & 0.03 & 
0.10 & 0.29  \\
& & 0.004 & 0.03 & 0.07 & 0.19 & 0.004 & 0.04 & 0.12 & 0.32 \\
\end{tabular}
\end{center} 
\end{table} 
\vskip 0.4cm

\vskip 0.4cm
\begin{table}[ht]
{\small Table V. Same as Table IV except for $\ov B_s\to VV$ decays.}
\begin{center}
\begin{tabular}{l c c c c c c c c c } 
 &  & \multicolumn{4}{c}{$\nc(LL)=2$}  
 &   \multicolumn{4}{c}{$\nc(LL)=\nc(LR)$}  \\ \cline{3-6} \cline{7-10}
\raisebox{2.0ex}[0cm][0cm]{Decay} & \raisebox{2.0ex}[0cm][0cm]{Class} & 
2 & 3 & 5 & $\infty$ & 2 & 3 & 5
& $\infty$  \\ 
\hline 
$\ov B_s\to K^{+*}\rho^-$ & I & 12.5 & 12.5 & 12.5 & 12.5 & 12.5 & 13.9
& 15.0 & 16.9 \\
& & 14.4 & 14.4 & 14.4 & 14.4 & 14.4 & 16.0 & 17.4 & 19.5 \\
$\ov B_s\to K^{0*}\rho^0$ & II & 0.40 & 0.40 & 0.40 & 0.40 & 0.40 & 0.044
& 0.094 & 0.72 \\
& & 0.46 & 0.46 & 0.46 & 0.46 & 0.46 & 0.051 & 0.11 & 0.84 \\
$\ov B_s\to K^{0*}\omega$ & II,VI & 0.26 & 0.21 & 0.19 & 0.17 & 0.26 & 0.04
& 0.02 & 0.28 \\
& & 0.30 & 0.25 & 0.22 & 0.19 & 0.30 & 0.044 & 0.031 & 0.32 \\
$\ov B_s\to K^{+*}K^{-*}$ & IV & 2.53 & 2.53 & 2.53 & 2.53 & 2.53 & 2.80
& 3.03 & 3.38 \\
& & 2.91 & 2.91 & 2.91 & 2.91 & 2.91 & 3.22 & 3.48 & 3.88 \\
$\ov B_s\to K^{0*}\ov K^{0*}$ & IV & 2.44 & 2.44 & 2.44 & 2.44 & 2.44 & 3.09
& 3.66 & 4.62 \\
& & 2.80 & 2.80 & 2.80 & 2.80 & 2.80 & 3.55 & 4.21 & 5.30 \\
$\ov B_s\to \rho^0\phi$ & V & 0.17 & 0.18 & 0.18 & 0.18 & 0.17 & 0.22
& 0.28 & 0.41 \\
& & 0.33 & 0.34 & 0.34 & 0.35 & 0.33 & 0.42 & 0.53 & 0.79 \\
$\ov B_s\to\omega\phi$ & V & 0.65 & 0.15 & 0.01 & 0.25 & 0.65 & 0.004
& 0.30 & 2.09 \\
& & 1.22 & 0.27 & 0.02 & 0.48 & 1.22 & 0.007 & 0.56 & 3.92 \\
$\ov B_s\to K^{0*}\phi$ & VI & 0.007 & 0.049 & 0.10 & 0.19 & 0.007 & 0.13
& 0.28 & 0.57 \\
& & 0.014 & 0.098 & 0.17 & 0.30 & 0.014 & 0.22 & 0.43 & 0.86 \\
$\ov B_s\to\phi\phi$ & VI & 13.8 & 8.77 & 5.57 & 2.15 & 13.8 & 7.15
& 3.40 & 0.37 \\
& & 25.1 & 15.9 & 10.1 & 3.91 & 25.1 & 13.0 & 6.18 & 0.68 \\
\end{tabular}
\end{center} 
\end{table} 
\vskip 0.4cm

\vskip 0.4cm
\begin{table}[ht]
{\small Table VI. Fractions of non-electroweak penguin contributions to
the branching ratios of penguin-dominated two-body $B_s$ decays, as defined by
Eq.~(3.8). Predictions are for $k^2=m_b^2/2$, $\eta=0.34,~\rho=0.16$, and 
$\nc(LR)=2,3,5,\infty$ with $\nc(LL)$ being fixed to be 2 in the first 
case and treated to be the same as $\nc(LR)$ in the second case. We use
the BSW model for form factors.}
\begin{center}
\begin{tabular}{l c c c c c c c c } 
 &  \multicolumn{4}{c}{$\nc(LL)=2$}  
 &   \multicolumn{4}{c}{$\nc(LL)=\nc(LR)$}  \\ \cline{2-5} \cline{6-9}
\raisebox{2.0ex}[0cm][0cm]{Decay} & 2 & 3 & 5 & $\infty$ & 2 & 3 & 5
& $\infty$  \\ 
\hline 
$\ov B_s\to \pi^0\eta'$ & 0.15 & 0.15 & 0.16 & 0.16 & 0.15 & 0.007 & 0.01
& 0.11  \\
$\ov B_s\to \pi^0\eta$ & 0.15 & 0.15 & 0.16 & 0.16 & 0.15 & 0.007 & 0.01
& 0.11  \\
$\ov B_s\to \pi^0\phi$ & 0.15 & 0.15 & 0.16 & 0.16 & 0.15 & 0.007 & 0.01
& 0.11  \\
$\ov B_s\to \rho^0\eta'$ & 0.15 & 0.15 & 0.15 & 0.14 & 0.15 & 0.007 & 0.01
& 0.11  \\
$\ov B_s\to \rho^0\eta$ & 0.15 & 0.15 & 0.15 & 0.14 & 0.15 & 0.007 & 0.01
& 0.11  \\
$\ov B_s\to \omega\eta'$ & 0.78 & 0.57 & 1.63 & 1.43 & 0.78 & 0.79 & 1.42
& 1.16 \\
$\ov B_s\to \omega\eta$ & 0.78 & 0.57 & 1.63 & 1.43 & 0.78 & 0.79 & 1.42
& 1.16 \\
$\ov B_s\to \phi\eta'$ & 1.73 & 1.70 & 1.69 & 1.65 & 1.73 & 1.93 & 0.63
& 0.61 \\
$\ov B_s\to \phi\eta$ & 1.71 & 2.16 & 3.01 & 0.39 & 1.71 & 2.00 & 2.58
& 2.81 \\
$\ov B_s\to K^0\phi$ & 3.25 & 0.23 & 0.49 & 0.07 & 3.25 & 0.43 & 0.68 &
0.82 \\
$\ov B_s\to K^{*+}K^{*-}$ & 0.87 & 0.87 & 0.87 & 0.87 & 0.87 & 0.94 & 1.00
& 1.08 \\
$\ov B_s\to K^{*0}\ov K^{*0}$ & 1.08 & 1.08 & 1.08 & 1.08 & 1.08 & 1.03 &
1.00 & 0.96 \\
$\ov B_s\to \rho^0\phi$ & 0.15 & 0.14 & 0.14 & 0.14 & 0.15 & 0.006 & 0.01
& 0.11 \\
$\ov B_s\to\omega\phi$ & 0.78 & 0.57 & 1.63 & 1.43 & 0.78 & 0.79 & 1.42 &
1.16 \\
$\ov B_s\to K^{0*}\phi$ & 3.82 & 0.55 & 0.76 & 0.86 & 3.82 & 0.75 & 0.84 &
0.87 \\
$\ov B_s\to\phi\phi$ & 1.25 & 1.32 & 1.41 & 1.69 & 1.25 & 1.32 & 1.43 &
2.19  \\
\end{tabular}
\end{center} 
\end{table} 
\vskip 0.4cm

\end{document}